\documentclass[aps,prx,superscriptaddress,twocolumn,longbibliography]{revtex4-1}
\usepackage[english]{babel}
\usepackage{amsmath}
\usepackage{graphicx}
\usepackage{color}

\newcommand{\rmc}{{\rm c}}
\newcommand{\rme}{{\rm e}}
\newcommand{\rmd}{{\rm d}}
\newcommand{\rmi}{{\rm i}}

\newcommand{\rmq}{{\rm q}}

\begin{document}

\title{Photon-mediated interactions: a scalable tool \\ to create and sustain entangled states of $N$ atoms}
\author{Camille Aron}
 \altaffiliation[Present address:]{ Laboratoire de Physique Th\'eorique, \'Ecole
Normale Sup\'erieure, CNRS, Paris, France. Instituut voor
Theoretische Fysica, KU Leuven, Belgium.}
\affiliation{Department of Electrical Engineering, Princeton University, Princeton, NJ 08544, USA}
\author{Manas Kulkarni}
\affiliation{Department of Physics, New York City College of
Technology, The City University of New York, Brooklyn, NY 11201, USA}
\author{Hakan E. T\"{u}reci}
\affiliation{Department of Electrical Engineering, Princeton University, Princeton, NJ 08544, USA}

\begin{abstract}
We propose and study the use of photon-mediated interactions for the generation of long-range steady-state entanglement between $N$ atoms. Through the judicious use of coherent drives and the placement of the atoms in a network of Cavity QED systems, a balance between their unitary and dissipative dynamics can be precisely engineered to stabilize a long-range correlated state of qubits in the steady state. We discuss the general theory behind such a scheme, and present an example of how it can be used to drive a register of $N$ atoms to a generalized W state and how the entanglement can be sustained indefinitely. The achievable steady-state fidelities for entanglement and its scaling with the number of qubits are discussed for presently existing superconducting quantum circuits. While the protocol is primarily discussed for a superconducting circuit architecture, it is ideally realized in any Cavity QED platform that permits controllable delivery of coherent electromagnetic radiation to specified locations.
\end{abstract}

\date{\today}
 
\maketitle

\section{Introduction}
\label{sec:intro}

Photon-mediated interactions are ubiquitous in nature. While the traditional formulation of quantum electrodynamics places equal emphasis on fields and sources, it is possible to take a point of view where the electromagnetic degrees of freedom are integrated out to reach an effective nonlocal field theory for matter only~\cite{feynman_mathematical_1950}. A particularly beautiful example of this point of view, most closely related to the phenomena investigated here, is Schwinger's formulation of the Casimir effect~\cite{schwinger_casimir_1975}. Here, the electromagnetic degrees of freedom are integrated out and result in a photon-mediated retarded and nonlocal interaction between two conducting surfaces.  

A classic example where photon-mediated interactions are at play is the superradiance (and subradiance) of a cluster of dipoles, first pointed out by Dicke~\cite{dicke_coherence_1954}. Here, photon-mediated interactions ultimately lead to generation of transient coherence between dipoles \cite{gross_superradiance_1982, brandes_coherent_2005}, which results in the emission of a powerful pulse whose intensity scales with $N^2$, where $N$ is the number of dipoles within a volume $V \sim \lambda^3$ ($\lambda$ is the wavelength of radiation). 
In free space, however, such interactions decay fast with interdipole distance, and it is challenging to engineer such interactions in a controlled way~\cite{devoe_observation_1996, eschner_light_2001}. Photon-mediated interactions are also at play in the self-organization transition of optically driven cold atoms in a cavity~\cite{ritsch_cold_2013}. In these systems, cavity-mediated long-range interactions between atoms, tunable by the drive strength, lead to softening of a motional excitation mode recently observed in experiments~\cite{baumann_dicke_2010, mottl_roton-type_2012}. Certain aspects of the underlying critical behavior of this non-equilibrium phase transition can be described through photon-mediated interactions between the atoms constituting the condensate~\cite{brennecke_real-time_2013, strack_dicke_2011, kulkarni_cavity-mediated_2013, konya_damping_2014}. 

In recent years, we have seen the first attempts to use such photon-mediated interactions to generate strong coupling and possibly entanglement between artificial atoms in solid-state cavity QED systems. These approaches capitalize on strong light-matter interactions that can be generated in confined geometries such as resonators~\cite{majer_coupling_2007, filipp_multimode_2011} and waveguides~\cite{loo_photon-mediated_2013}. In particular, Ref.~\cite{loo_photon-mediated_2013} demonstrated coherent exchange interactions between two superconducting qubits separated by as much as a full wavelength (in that case $\lambda \sim 18.6$ mm) in an open quasi-1D transmission line. More recently, superradiance of two artificial atoms was observed and characterized in a controlled setting in a superconducting quantum circuit \cite{mlynek_observation_2014}.

\begin{figure}[!t]
\includegraphics[width=8.5cm]{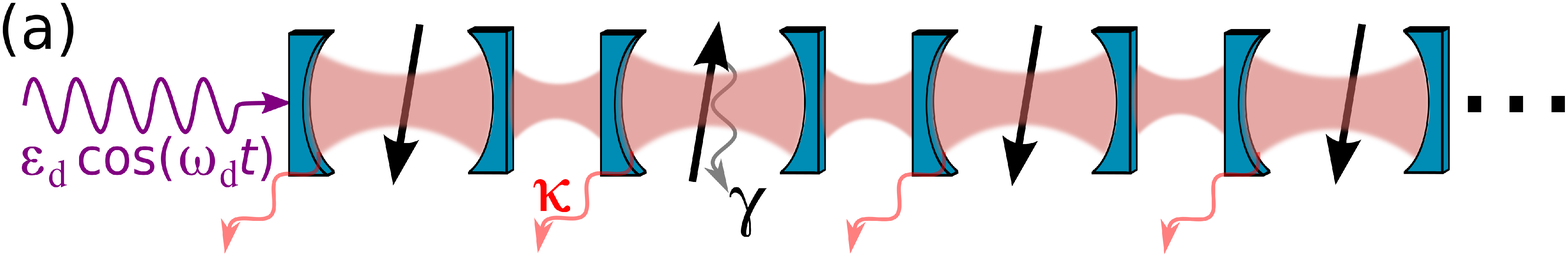}
\vspace{1pt} 
\includegraphics[width=7.0cm]{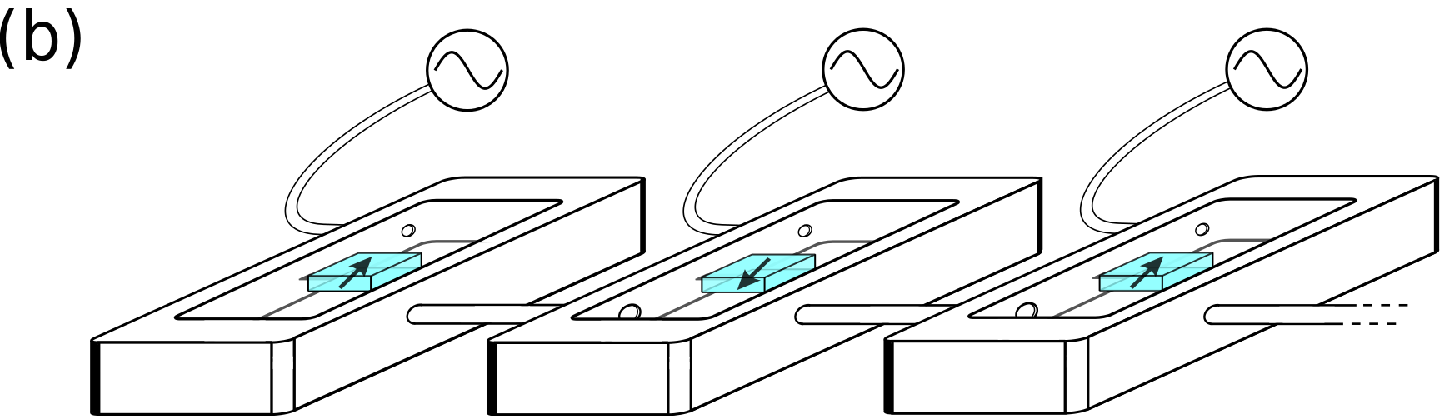} 
\caption{\footnotesize (color online) (a) One-dimensional array of cavity-qubit systems coupled by photon exchange and subject to one or several ac microwave drives, cavity decay at rate $\kappa$, qubit relaxation $\gamma$ and pure dephasing $\gamma_\phi$.
(b) Implementation with superconducting transmon qubits embedded in interconnected microwave cavities and driven by external continuous-wave generators.}
\label{figschem1d}
\end{figure}

The goal of this paper is to show how photon-mediated interactions between qubits embedded in an engineered electromagnetic environment can be harnessed to controllably generate a certain large-scale entangled state of $N$ qubits {\it in the steady state}. In the present work, the role of photons is twofold. First, they mediate a coherent coupling between distant qubits. Second, they provide a controllable dissipative mechanism which can be used to stabilize a long-range correlated many-body state of the qubits. By driving the system at a suitable frequency, one can achieve a transition that produces the desired many-body state while dumping energy into one of the electromagnetic modes. The dissipative mechanisms are key to make the scheme steady-state, in contrast to better-known alternatives, such as those based on Rabi cycling between a ground state and a desired excited state. We show how the balance of the unitary and dissipative contributions can be precisely tuned by the placement of qubits in an engineered photonic environment, and a set of coherent electromagnetic drives with specified amplitudes and frequencies. 

Engineered dissipative dynamics has recently been employed in superconducting circuits to cool a single qubit to a desired state on the Bloch sphere~\cite{murch_cavity-assisted_2012}, and two qubits that reside in a single cavity to a Bell state~\cite{shankar_autonomously_2013}. Furthermore, a number of recent theoretical works have focused on generation of high-fidelity steady-state entanglement between two superconducting qubits~\cite{leghtas_stabilizing_2013,reiter_steady-state_2013,huang_generation_2013,aron_steady-state_2014}. We present the general theory underlying these phenomena, deriving and solving the dynamics of $N$ qubits residing in an arbitrary open electromagnetic environment, subject to coherent driving and losses. {The underlying new principle is based on using collective electromagnetic (EM) modes of a large structure (such as Bloch modes of a lattice of cavities) to dissipatively stabilize a collective state of spins. By ``scalability'' we refer to the fact that, for the specific protocol worked out here for W states, that the same protocol used for $N$ qubits can be used for $N+1$ qubits. Generally it is to be expected that the fidelity of stabilization degrades as $N$ is increased because of spectral crowding, which we discuss for the particular protocol we investigate here, and provide an estimate for the maximum number of qubits that can be stabilized reliably. Interestingly, spatial symmetries in the system (such as discrete translational invariance) can be used along with a {\it spatially} modulated coherent drive, to greatly enhance the spectral resolution. In one of the protocols we propose, this allows us to resolve W states in a way that the state selectivity is limited by only the finesse of the EM system, instead of the mean level spacing of the collective spin states.}

In Sec.~\ref{sec:1dmodel}, we lay out a particular architecture that we have in mind for a proof-of-principle demonstration: we consider of model of $N$ qubits residing in a one-dimensional array of cavities. We present the different layers of approximations that allow us to have an analytic handle over the problem. First, we show how cavity photons mediate effective qubit-qubit interactions through the collective electromagnetic modes of the array. In Sec.~\ref{sec:cooling}, we show how the {\it full Liouvillian} describing the evolution of the reduced density matrix of the $N$ qubits can be engineered to drive the qubit subsystem to generalized W states, and the entanglement sustained as long as the drives are on. The fabrication tolerances of such a cavity array has recently been studied experimentally~\cite{underwood_low-disorder_2012}. Thus, the controlled fabrication of such arrays is well within reach of the current superconducting circuit technology both in coplanar and 3D configurations~\cite{underwood_thesis_2015}. We present the fidelities that can be expected in recently fabricated systems and analyze the fault tolerance of the method to phase and amplitude noise of the drive parameters, as well as the nonuniformity of qubit and cavity parameters. Finally, in  Sec.~\ref{sec:ph-mediated}, we generalize our scheme to arbitrary arrangements of qubits coupled to an engineered photonic backbone, going beyond the tight-binding approximation for the EM system used in previous sections.

\section{One-dimensional lattice model}\label{sec:1dmodel}

Consider a one-dimensional array of $N$ identical microwave cavities with nominally equal frequencies $\omega_\rmc$, coupled to each other capacitively described by a nearest-neighbor tunneling matrix element $J$. Each cavity houses a superconducting qubit with splitting $\omega_\rmq$. We shall consider both the cases of open (\textit{i.e.} non-periodic) and periodic boundary conditions (identifying site $N$ with site 0). Later, we will relax our assumptions on identical cavity resonance frequencies and consider the most general case, showing that the general approach to the dissipative stabilization of a generalized W state stands. 

Furthermore, each cavity shall be driven by a coherent monochromatic microwave source with frequency $\omega_\rmd$ and a site-dependent amplitude $\epsilon^{\rmd}_i$ and phase $\Phi_i$; see Fig.~\ref{figschem1d}(b) for a 3D superconducting circuit architecture of the system Fig.~\ref{figschem1d}(a). We work in a regime where $\omega_\rmc$, $\omega_\rmq$ and $\omega_\rmd$ are mutually far detuned from each other (typically on the order of GHz).
The starting Hamiltonian is that of a one-dimensional driven Jaynes-Cummings lattice model studied before in Refs.~\cite{knap_emission_2011,nissen_nonequilibrium_2012,grujic_non-equilibrium_2012,hur_many-body_2015},
\begin{align} \label{eq:H}
 H(t) = H_\sigma + H_{\sigma a} + H_a(t) \,,
 \end{align}
where $H_\sigma$,  $H_{\sigma a}$, and $H_a(t)$ are respectively the qubit, the Jaynes-Cummings light-matter coupling, and the driven cavity Hamiltonians,
 \begin{align}
H_\sigma =& \! \sum_i   \omega_\rmq \frac{\sigma_i^z}{2},\,    H_{\sigma a} =  g \! \sum_i\! \left[ a_i^\dagger \sigma^-_i + \mbox{H.c.} \right]  \,,
\\
H_a(t) =& \! \sum_i\!  \left[ \omega_\rmc a_i^\dagger a_i 
   - J ( a_{i}^\dagger a_{i+1} + \mbox{H.c.} ) \right.
      \nonumber \\
  &   \qquad \left. +  2 \epsilon^{\rmd}_i \cos(\omega_\rmd t + \Phi_i) \left( a_i + a^\dagger_i \right)\right] \,. \label{eq:Ha}
\end{align}
Here, $i$ runs from site $0$ to $N-1$. 
The qubits are two-level systems described by the usual Pauli pseudo-spin operators $\sigma_i^{x,y,z}$ and $\sigma_i^\pm \equiv (\sigma^x_i \pm \rmi\sigma_i^y)/2$.
$\epsilon^\rmd_i$ and $\Phi_i$ are, respectively, the cavity-dependent amplitude and phase of the ac microwave drives.
Thermal equilibrium is achieved by setting all drive amplitudes to zero, $\epsilon^\rmd_i=0$. Without loss of generality, we assume that the detuning between cavity and qubit frequency  $\Delta \equiv \omega_\rmq - \omega_\rmc > 0$, and $J>0$. We operate in the dispersive regime corresponding to $g/\Delta \sim 10^{-1}$ 
and at sufficiently weak drive amplitudes to ensure the presence of very few photons in the cavities, so the Schrieffer-Wolff perturbation theory to be applied shortly is well justified. Below, we entirely integrate out the photonic degrees of freedom, resulting in (a) an effective qubit-qubit interaction as discussed on more general grounds in Sec.~\ref{sec:ph-mediated}, (b) local Zeeman fields of the form $\epsilon^\rmd_i \sigma_i^{x}$ (for $\Phi_i=0$), and (c) a non-unitary evolution characterized by controllable transition rates.

\paragraph*{Intrinsic dissipation in the system.}
In addition to the unitary dynamics described by Eq.~(\ref{eq:Ha}), we assume the individual qubits are coupled to uncontrolled environmental degrees of freedom that give rise to single qubit spin-flip rate ($\gamma$), a single-qubit pure dephasing rate $\gamma_\phi$, and a cavity decay rate $\kappa$. 

\paragraph*{Typical system parameters.} In a recent experiment studying the $N=2$ case of the protocol described here \cite{ucbpaper} in a 3D superconducting circuit architecture, typical system parameters were:  $\omega_\rmc \simeq 7, \omega_\rmq \simeq 6, g \simeq J \simeq 10^{-1}, \kappa \simeq 7 \times 10^{-4}, \gamma \simeq 4 \times 10^{-5}$ all in units of $2\pi$~GHz. Below, our analytical approach is performed assuming the hierarchy of energy scales $\Delta \gg g, J \gg \kappa \gg \gamma \gg \gamma_\phi$.


\paragraph*{Rotating wave approximation.}
We eliminate the time dependence in $H_a(t)$ by working in the frame rotating at $\omega_\rmd$ and dropping the non-secular terms. In the rest of the Hamiltonian (\ref{eq:H}), this also amounts to replacing $\omega_\rmq$ by $\Delta_\rmq\equiv \omega_\rmq - \omega_\rmd$ and $\omega_\rmc$ by $-\Delta_\rmc\equiv \omega_\rmc - \omega_\rmd$. 
Once expressed in the eigenbasis of the (undriven) coupled cavity system, $H_{a}$ is given by
\begin{align} \label{eq:Hak}
  H_{a} = \sum_k (\omega_k -\omega_\rmd) a^\dagger_k a_k  +  ( \epsilon^\rmd_k a^\dagger_k + \mbox{H.c.} )\,.
\end{align}
Here, $k$ is the discrete Bloch wavevector, $a_k^\dagger = \sum_j \varphi_{k}^*(j) \, a_j^\dagger$ creates a photon in the $k^{\mathrm{th}}$ mode. The specific discrete values of $k$ and the corresponding mode profiles $\varphi_k(j)$ are to be fixed by the boundary conditions. For periodic boundary conditions, the set of quasi-momenta are $k= 2\pi\, n/N$, with $n = 0,\ldots, N-1$ and $\varphi_{k}(j) = \rme^{-\rmi k j} / \sqrt{N}$. 
For open boundary conditions, $ k = \pi (n+1)/(N+1)$ and $\varphi_k(j) = \sqrt{2/(N+1)} \sin\left(k(j+1)\right)$. In Eq.~(\ref{eq:Hak}), we incorporated the phase $\Phi_j$ in $\epsilon^\rmd_j$, which can henceforth be complex and  $\epsilon^\rmd_k =   \sum_j \varphi_{k}(j) \, \epsilon_j^\rmd $.
 The eigenfrequencies are given by the photonic dispersion relation
\begin{align}
\omega_k = \omega_\rmc - 2 J \cos(k)\,.
\end{align}

\begin{figure}[!t]
\includegraphics[width=3.5cm, angle=0]{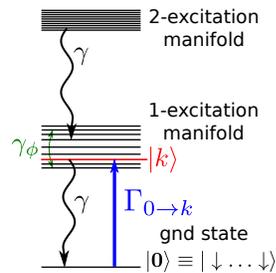} 
\caption{\footnotesize (color online)
Engineered qubit many-body  spectrum. The non-equilibrium drive and the resulting photon fluctuation bath are used to create a dominant transition from the trivial ground state to a target entangled W state in the 1-excitation manifold. The existence of a non-zero spontaneous decay $\gamma$ is critical to avoid populating higher-excited state manifolds.} \label{fig:mediated}
\end{figure}

\paragraph*{Effective dissipative $XY$~model.}
We eliminate the light-matter interaction to second-order perturbation theory in $g/\Delta$ by means of a Schrieffer-Wolf (SW) transformation which maps $H \mapsto \rme^{X} H  \rme^{X^\dagger}$ where
\begin{align}
X \equiv g \sum_k \left[ \frac{a_k \sigma_k^\dagger}{\omega_\rmq - \omega_k}
-\mathrm{H.c.}
 \right]\,.
\end{align}
Collecting all the terms, we obtain an isotropic $XY$ model subject to a magnetic field for the qubit subsystem ($H_\sigma$), weakly coupled to the photon fluctuations of the EM backbone ($H_{\sigma a}$):
 \begin{align}
H_\sigma =& \! \sum_i  
\boldsymbol{h_i} \cdot \frac{\boldsymbol{\sigma}_i}{2} - \frac{J}{2} \left(\frac{g}{\Delta}\right)^2 \left[ \sigma^x_{i} \sigma^x_{i+1} + \sigma^y_{i} \sigma^y_{i+1}\right]\,, \label{eq:Hspinchain} \\
H_{\sigma a} &= \sum_i  \left(\frac{g}{\Delta}\right)^2 \sigma_i^z( \Delta a_i^\dagger a_i + \epsilon^\rmd_i a_i^\dagger +{\epsilon^\rmd_i}^* a_i)\,.
\label{eqtransXY}
\end{align}
Here, ${h}^x_i = 2 \mbox{Re}(\epsilon^\rmd_i) (g/\Delta)$, $h^y_i= -2 \mbox{Im}(\epsilon^\rmd_i) (g/\Delta)$, $h^z_i  = \Delta_\rmq + \Delta (g/\Delta)^2 $.
 The effective magnetic field $\boldsymbol{h}_i$ is mainly oriented along the $z$ direction but we shall see that, while they break the integrability of the model, the $x$ and $y$ components of the emergent Zeeman field play a crucial role for the scheme below.

We note that the SW transformation is also responsible for subleading corrections to the strength of the dissipative terms (\textit{e.g.} the qubit decay $\gamma$ acquires a contribution from the cavity decay $\kappa$, corresponding to a Purcell contribution, and \textit{vice versa}) the local contributions of which can be simply included by working with renormalized parameters $\kappa$, $\gamma$ and $\gamma_\phi$ at low photon occupations~\cite{boissonneault_dispersive_2009}, which is the regime studied here. These effective parameters can be extracted from experimental measurements. 

When the drives are off, $\epsilon^\rmd_i = 0$, the system simply thermalizes with its environment -- this is just the physics of blackbody radiation in a coupled cavity system (including $N$ dipoles). Viewed from the perspective of the qubits, while $H_\sigma$ describes a quantum phase transition from a paramagnetic to a ferromagnetic phase when the magnitude of the transverse field is on the order of the nearest-neighbor coupling $J ( g/\Delta)^2$, this regime is never reached for realistic system parameters. 
For Raman-driven qubits the situation is more interesting, and the phase diagram was recently studied displaying various exotic attractors~\cite{schiro_exotic_2015}.
Henceforth, we only work in the experimentally achievable regime $\omega_\rmq \gg  J(g/\Delta)^2$  and, to leading order, the ground state of $H_\sigma$ is simply the separable state $|\boldsymbol{0} \rangle  \equiv | \! \downarrow \ldots \downarrow \rangle$. 
The low-energy spectral content of the qubit sector is depicted in Fig.~\ref{fig:mediated}: the first $N$-qubit excited manifolds are roughly separated by $\Delta_\rmq$ while the lifting of degeneracy within each manifold is controlled by $J (g/\Delta)^2$

From the point of view of equilibrium statistical mechanics, the many-body system of qubits described by $H_\sigma$ is not interesting because it would thermalize with the radiative reservoir, reaching a steady state $\rho_\sigma (t \rightarrow \infty) \propto \rme^{-\beta H_\sigma} \sim | \boldsymbol{0}  \rangle \langle \boldsymbol{0}|$, a collection of uncorrelated qubits in their ground state. 
We shall see that turning on the drives will change this situation dramatically. A careful choice of drive parameters (even for small amplitudes of drives) will be shown to enable the stabilization of a particular many-body state of qubits in the excited-state manifold. 

\section{Stabilization of generalized W states}\label{sec:cooling}
Our goal in this section is to identify non-trivial entangled eigenstates of the spin chain $H_\sigma$ and design a protocol which, starting from the ground state $|\boldsymbol{0}\rangle$ that can be straightforwardly prepared, achieves the stabilization of an interesting excited state of choice. Below we discuss the details of a robust protocol for the stabilization of a generalized W state of qubits with {\it minimal} resources.

\paragraph*{Linearized photon spectrum.}
We first address the non-linearities in $H_{\sigma a}$ by decomposing the photonic field into mean-field plus bosonic fluctuations:
\begin{align}
a_k \equiv \bar a_k + d_k \mbox{ with }  \bar a_k = \frac{\epsilon^\rmd_k}{\omega_\rmd - \omega_k + \rmi \kappa/2}\;. 
\label{eqsmallfluct}
\end{align}
We assume here that all Bloch modes have the same loss $\kappa$. This can be made more precise but it will not qualitatively change the results we present. 
Neglecting those terms that are quadratic in the fluctuations and that couple to the qubits, the light sector reduces to 
\begin{align}
H_a \rightarrow H_d &= \sum_k (\omega_k - \omega_\rmd) d_k^\dagger d_k \,, \label{eq:bath1} \\
H_{\sigma a} \rightarrow H_{\sigma d} &= \left(\frac{g}{\Delta}\right)^2  \sum_i  \sigma^z_i [( \Delta \bar{a}_i  + \epsilon^\rmd_i ) d_i^\dagger + \mbox{H.c.} ]\,. \label{eq:bath2}
\end{align}

\paragraph*{Diagonalization of the matter sector.}
The Hamiltonian Eq.~(\ref{eq:Hspinchain}) in the presence of non-zero drive terms is non-integrable. We therefore proceed by projecting it into the low-energy sector with a maximum one excitation:
\begin{align} \label{eq:Hstrunc}
H_\sigma  =\sum_k E_k |k \rangle \langle k | + \left( \frac{g}{\Delta} \right) \left( \epsilon^\rmd_k  |k\rangle\langle \boldsymbol{0} | + \mbox{H.c.} \right),
\end{align} 
We have set the energy of the ground state $|\boldsymbol{0} \rangle  \equiv | \downarrow \ldots \downarrow \rangle$ to zero ($E_{\boldsymbol{0}} = 0$). Here the states 
\begin{equation}
|k\rangle = \sum_{i=0}^{N-1}  \varphi_k^* (i) \ |i\rangle\,,
\end{equation}
with $|i\rangle \equiv |\downarrow_0 \ldots\downarrow_{i-1} \, \uparrow_i \, \downarrow_{i+1} \ldots \downarrow_{N-1}  \rangle$ indicating one excitation located at site $i$, are states that carry a single qubit excitation of quasi-momentum $k$, entangled over the entire chain. These are the eigenstates of the {\it undriven} spin chain (\textit{i.e.} for $\epsilon^\rmd_i =0 \, \forall\, i$) with a dispersion relation
\begin{align}
E_k = \epsilon_k - \omega_\rmd, \; \epsilon_k = \omega_\rmq + \delta\omega_\rmq - 2 J \left(\frac{g}{\Delta}\right)^2 \! \cos(k)\;.
\end{align} 
Here $\delta\omega_\rmq \simeq  g^2/\Delta$  is the cavity-induced Stark shift. This truncation of the Hamiltonian holds if the higher-excitation manifolds are not significantly occupied during the dynamics. This can be checked a posteriori and we do so. 

Let us first discuss the case of open boundary conditions for which the absence of translational and space-reversal symmetry generically yields a fully non-degenerate spectrum. The effect of the drive term in Eq.~(\ref{eq:Hstrunc}), assumed to be small as stated before, can be taken into account through a perturbation theory and yields the following eigenstates of $H_\sigma$ to lowest order in $({g}/{\Delta}) (\epsilon_k^\rmd/\Delta_\rmq)$:
\begin{align}
|\widetilde{\boldsymbol{0}} \rangle 
 &\simeq
  |\boldsymbol{0} \rangle \!-\! \left(\frac{g}{\Delta}\right) \!  \sum_{k}  \! \frac{\epsilon^\rmd_k}{\Delta_\rmq} |k \rangle,\, 
\widetilde{E}_{\boldsymbol{0}} \simeq \!-\!  \left(\frac{g}{\Delta}\right)^2 \! \sum_k \! \frac{|\epsilon^\rmd_k|^2}{\Delta_\rmq},\\
|\widetilde{k} \rangle & \simeq
|k \rangle + \left(\frac{g}{\Delta}\right)  \frac{{\epsilon^\rmd_k}^*}{\Delta_\rmq} |\boldsymbol{0} \rangle,\,
\widetilde{E}_k \simeq  E_k + \left(\frac{g}{\Delta}\right)^2 \frac{|\epsilon^\rmd_k|^2}{\Delta_\rmq}.
\label{eqqbitcollen}
\end{align}
The above corrections to the undriven eigenstates are crucial for the success of the two-photon cooling mechanism presented below.

\paragraph*{Transition rates.}

By virtue of Eq.~(\ref{eqsmallfluct}), the coupling of the photonic fluctuations (on top of a classical part) to the spin-chain Eq.~(\ref{eqtransXY}) can be treated in perturbation theory. This permits us to integrate them out arriving at an effective master equation for qubits only. We note that the fluctuations of the collective photon modes of the lattice, described by spectral function per mode $q$
\begin{align}
\rho_q(\omega) = - \frac{1}{\pi} \mbox{Im } \frac{1}{\omega -\omega_q+\rmi\kappa/2}\;,
\end{align}
can be easily manipulated by the design of the cavity lattice. Assuming that the photon fluctuations, which couple to the spin degrees of freedom via {Eq.~(\ref{eq:bath2})}, thermalize with the radiative environment which in turn is taken to be at very low temperature, we arrive at the effective master equation for spins only $\rho_\sigma$ at steady state, $\rho_\sigma^{\rm NESS} \equiv \lim\limits_{t\to\infty} \rho_\sigma$:
\begin{align}\label{eq:master}
\partial_t \rho^{\rm NESS}_\sigma & \! \! =  0 =  -\rmi \left[ H_\sigma,\rho^{\rm NESS}_\sigma \right]
+ 
\sum_k \Gamma_{\boldsymbol{0}\to k} \mathcal{D}[| \widetilde{k} \rangle \langle \widetilde{\boldsymbol{0}} | ] \rho^{\rm NESS}_\sigma \nonumber \\
& \hspace{-3em} + \gamma \sum_k  \mathcal{D}[| \widetilde{\boldsymbol{0}} \rangle \langle \widetilde{k} | ]  \rho^{\rm NESS}_\sigma
 +  \frac{2\gamma_\phi}{N} \sum_{k\,q} \mathcal{D}[| \widetilde{q} \rangle \langle \widetilde{k} | ] \rho^{\rm NESS}_\sigma.
\end{align}
The derivation of this master equation relies on the separation of time scales: the relaxation time scale of cavity fluctuations $d_k$, on the order of $1/\kappa$, is much shorter than that of the reduced density matrix of the spins $\rho_\sigma(t)$. This separation of time scales is perfect in the steady state \cite{afoot,gsc,amitra} which we are interested in here.  

The Lindblad-type operators are defined as $\mathcal{D}[X] \rho  \equiv \left( X \rho X^\dagger - X^\dagger X \rho + \rm{H.c.} \right)/2$ and the $\Gamma_{\boldsymbol{0} \to k} $'s correspond to non-equilibrium transition rates  between the ground state $|\widetilde{\boldsymbol{0}} \rangle$ and a given excited many-body state $| \widetilde{k} \rangle$ [see Eq.~(\ref{eqqbitcollen})] in the one-excitation manifold of $H_\sigma$, given in Eq.~(\ref{eq:Hspinchain}). These are found to be 
\begin{align}\label{eq:rate}
\Gamma_{\boldsymbol{0} \to k} = 2 \pi \sum_q   \Lambda_{kq}^2\rho_q(\omega_\rmd+\widetilde{E}_{\boldsymbol{0}} -\widetilde{E}_k)\,,
\end{align}
where the transition matrix element is given by
\begin{align}
\Lambda_{kq} =& \left| 
 \left(1+2\frac{\Delta}{\Delta_\rmc} \right) \!\!
 \frac{1}{\Delta_\rmq} \!\! \left(\frac{g}{\Delta}\right)^3 \!\!\!
 \sum_{k' k'' q'} \!\!\! f_{k k' k''} f_{q' q k'}^* \epsilon^\rmd_{k''} \epsilon^\rmd_{q'} \right|\,,
\end{align}
with the tensor $f_{k k' k''} \equiv \sum_i \varphi_k(i) \varphi_{k'}^*(i) \varphi_{k''}^*(i)$. The integration over the photon-fluctuation degrees of freedom also yields Lamb-shift corrections of the energy levels, but this does not play any substantial role in our scheme, see discussion below Eq.~(\ref{eq:wdopt}).

\paragraph*{Dynamics.}
By virtue of having written the steady-state master equation Eq.~(\ref{eq:master}) in the eigenbasis of $H_{\sigma}$, all off-diagonal matrix elements of $\rho_\sigma^{\rm NESS}$ by construction vanish as the steady state is approached. Consequently, the dynamics can be faithfully described by effective rate equations for the populations of eigenstates, $n_{\boldsymbol{0}}$ and $n_k$:
\begin{align} 
\frac{\rmd n_{\boldsymbol{0}}}{\rmd t} &=  \gamma \, \sum_q n_q  - \Gamma_{\boldsymbol{0}\to q} \, n_{\boldsymbol{0}}  \label{eq:eom1} \\
\frac{\rmd n_k}{\rmd t} &=  - \gamma \, n_k  +   \Gamma_{\boldsymbol{0}\to k} \, n_{\boldsymbol{0}}
 + \frac{2 \gamma_\phi}{N} \sum_{q} (n_q -n_k\, ).  \label{eq:eom2}
\end{align}
The terms in $\gamma$ correspond to qubit decay, flipping down the pseudo spins and relaxing the energy by $\Delta_\rmq$ (in the rotating frame). The terms in $\gamma_\phi$ correspond to pure dephasing processes, the action of which is to equalize the populations of the states in the one-excitation manifold. The emergent level structure and rates are summarized in Fig.~\ref{fig:mediated}. 

We note that, while the full dynamical evolution of the qubit-EM system (viz. Eq.~(\ref{eq:H}) in the presence of qubit and cavity decay) is clearly non-Markovian, the proper secularization of the equations around the operation frequency $\omega_d$ allows us to describe the qubit dynamics through the relatively transparent rate equations~(\ref{eq:eom1} and \ref{eq:eom2}).

Irrespective of the initial conditions, Eqs.~(\ref{eq:eom1} and \ref{eq:eom2}) have a unique non-equilibrium steady-state solution and, after transient dynamics, the occupation of the state $|k\rangle$ is given by 
\begin{align}
n_k^{\rm NESS} &= \frac{1}{1+2\gamma_\phi/\gamma} \frac{\Gamma_{\boldsymbol{0} \to k} + (2\gamma_\phi/N\gamma) \sum_q \Gamma_{\boldsymbol{0} \to q} }{\gamma + \sum_q \Gamma_{\boldsymbol{0} \to q}} 
\,. \label{eq:nqness}
\end{align}

\paragraph*{Stabilization protocol.}
Equation~(\ref{eq:nqness}), together with Eq.~(\ref{eq:rate}), transparently elucidates how to stabilize a given pure entangled state of qubits $|k\rangle$ in the steady state. The protocol requires the maximization of $\Gamma_{\boldsymbol{0}\to k}$, given in Eq.~(\ref{eq:rate}), to make it the largest of all rates among ($\{ \Gamma_{\boldsymbol{0}\to q} \}$, $\gamma$, $\gamma_\phi$). This is performed by optimally tuning the drive frequency  $\omega_\rmd$ such that the sharply peaked photonic spectrum $\rho_q(\omega_\rmd+\widetilde{E}_{\boldsymbol{0}} -\widetilde{E}_k)$ in Eq.~(\ref{eq:rate}) reaches the maximum amplitude of the Lorentzian, which is on the order of $1/\kappa$. This is possible whenever there is at least one mode $q_0$ with $\Lambda_{k q_0} \neq 0$ and the optimum $\omega_\rmd$ is the solution of the energy-conservation equation $\omega_\rmd+\widetilde{E}_{\boldsymbol{0}} -\widetilde{E}_k = \omega_{q_0}$, \textit{i.e.} 
\begin{align}
\omega_\rmd = &  \frac{\omega_\rmq + \delta\omega_\rmq+\omega_\rmc}{2} - J \cos(q_0)  \nonumber \\
& \quad + \left(\frac{g}{\Delta}\right)^2 \left[ - J \cos(k) + \frac12 \sum_{q\neq k} \frac{|\epsilon^\rmd_q|^2}{\Delta_\rmq} \right]
\,.
\label{eq:wdopt}
\end{align}
This energy-matching condition describes a one-photon process in the rotating frame equivalent to a two-photon process in the laboratory frame. The corresponding Raman inelastic scattering process uses the energy of the two incoming drive photons to perform the qubit transition while simultaneously dumping a photon in one of the cavity modes \cite{ucbpaper}. We note that when Eq.~(\ref{eq:wdopt}) is satisfied, the Lamb-shift correction of $\widetilde{E}_q$ vanishes.

\begin{figure}[t!]
\includegraphics[width=8.4cm]{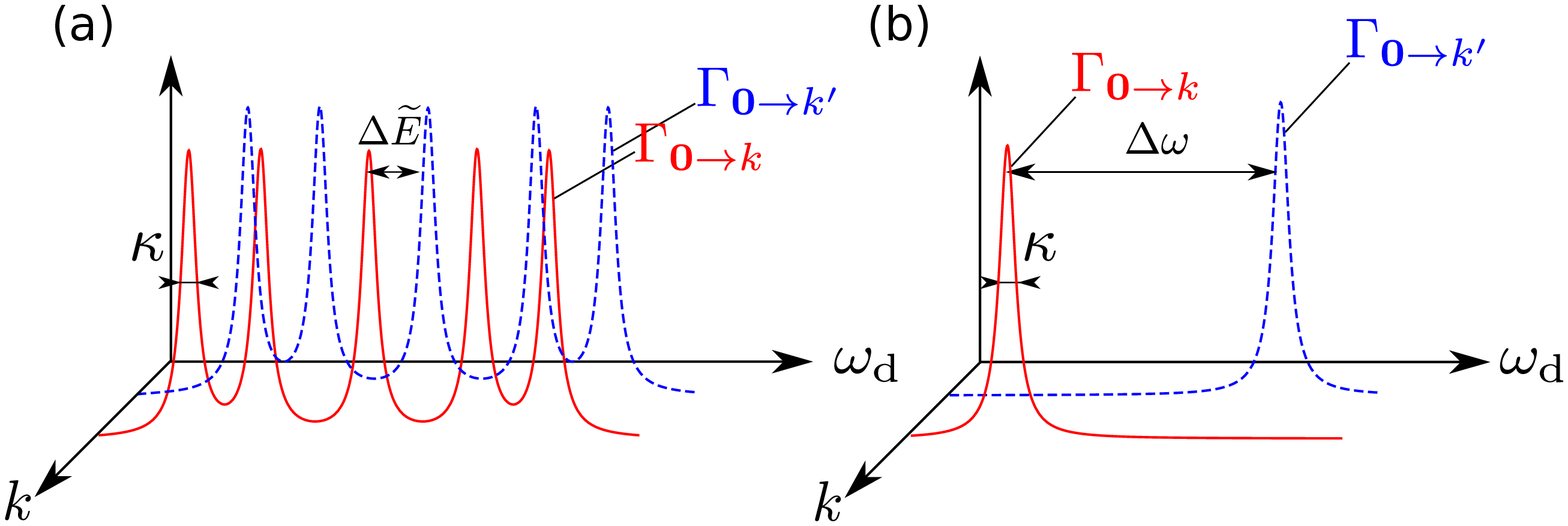} 
\caption{\footnotesize  \label{fig:gamma} (color online) Schematics of the pumping rate $\Gamma_{0 \to k}$ as a function of the drive frequency $\omega_\rmd$ for $N=5$. (a) Driving first cavity only: any of the five photon-fluctuation modes (responsible for the five peaks) can be used for dissipative stabilization.  
Driving to $|k\rangle$ while avoiding populating the nearest state $|k'\rangle$ 
necessitates $\kappa < \Delta \widetilde{E} \equiv |\widetilde{E}_k - \widetilde{E}_{k'}| \sim  2\pi J (g/\Delta)^2 /N$. 
(b) Driving all cavities equally: only one of the five modes can channel the mechanism and driving to the nearest state $|k'\rangle$ is avoided if  $\kappa < \Delta \omega \equiv |\omega_k - \omega_{k'}| \sim  2\pi J / N$. }
\end{figure}

\paragraph*{{Scalability and limitations.}}
Equation~(\ref{eq:nqness}) sets an upper bound on the fidelities, $n_{k}^{\rm NESS} \leq n^{\rm max} =
(\gamma+2\gamma_\phi/N)/(\gamma+ 2 \gamma_\phi)$, which highlights the necessity of working with qubits that have a pure dephasing rate $\gamma_\phi$ much smaller than their relaxation rate $\gamma$. This upper bound is not tight but allows us to highlight the role of the pure dephasing mechanisms. The success of the protocol, {and its scalability to large N}, also relies, via the numerator of Eq.~(\ref{eq:nqness}), on the resolving power of the spectral width of the photon density of states ($\sim \kappa$) \textit{i.e.} the precision with which the photon fluctuations can target the spin-chain state $|k\rangle$ without exciting other eigenstates close in energy. The limitations on the resolving power depend strongly on the drive spatial profile. This is illustrated here considering two extreme cases, one where only one cavity is driven ($\epsilon^\rmd_i = \epsilon^\rmd \delta_{i1}$), the other corresponding to a case where all cavities are driven with equal amplitude ($\epsilon^\rmd_i = \epsilon^\rmd \, \forall i$). In the first case [Fig.~\ref{fig:gamma}(a)], the transition rates for every $0 \to k$ have multiple peaks. Thus the stabilization of $|k\rangle$ at the optimal frequency given by Eq.~(\ref{eq:wdopt}) while avoiding the population of the nearest state $|k'\rangle$ requires the condition $\kappa < \Delta \widetilde{E} \equiv |\widetilde{E}_k - \widetilde{E}_{k'}| \sim  2\pi J (g/\Delta)^2 /N$.  On the other hand, the second case, driving each site identically yields rates that have a single peak, that for neighboring spin chain states $k$ and $k'$ are separated by the free spectral range (of the collective EM modes) of the cavity chain, $\Delta \omega \equiv |\omega_k - \omega_{k'}| \sim  2\pi J / N$.
{This indicates that the protocol with uniform driving can be scaled up to a number of qubits on the order of $N_{\rm max} \lesssim 2 \pi J / \kappa$.}

This situation is easily understood: the rates given in Eq.~(\ref{eq:rate}) contain the fulfillment of an energy-conservation condition, which can, in principle, be satisfied picking any collective EM mode $q$. The set of non-zero transition matrix elements in the sum Eq.~(\ref{eq:rate}) can, however, be substantially narrowed down by choosing a drive amplitude profile ($\epsilon^\rmd_{i}$) that is narrow in the momentum domain. For example, $\epsilon^\rmd_{p} = \delta_{pp_0}$, collapses the sum to a single term by imposing a quasi-momentum conservation condition between $p_0$, the target spin-chain state with momentum $k$, and $q_0$, the quasi-momentum of the collective EM mode picked for stabilization (note that conservation of momentum is strictly valid in periodic systems). This is given by $k = 2p_0 - q_0$ and is consistent with the interpretation of stabilization via a two-photon process. For such driving, the optimal frequency of the drive $\omega_\rmd$ is then given by Eq.~(\ref{eq:wdopt}). We note that this transparent criterion was used in Ref.~\cite{schwartz_toward_2015} to selectively stabilize either the triplet or the singlet state of two transmon qubits.

\begin{figure}[t!]
\includegraphics[width=6.5cm, angle=-90]{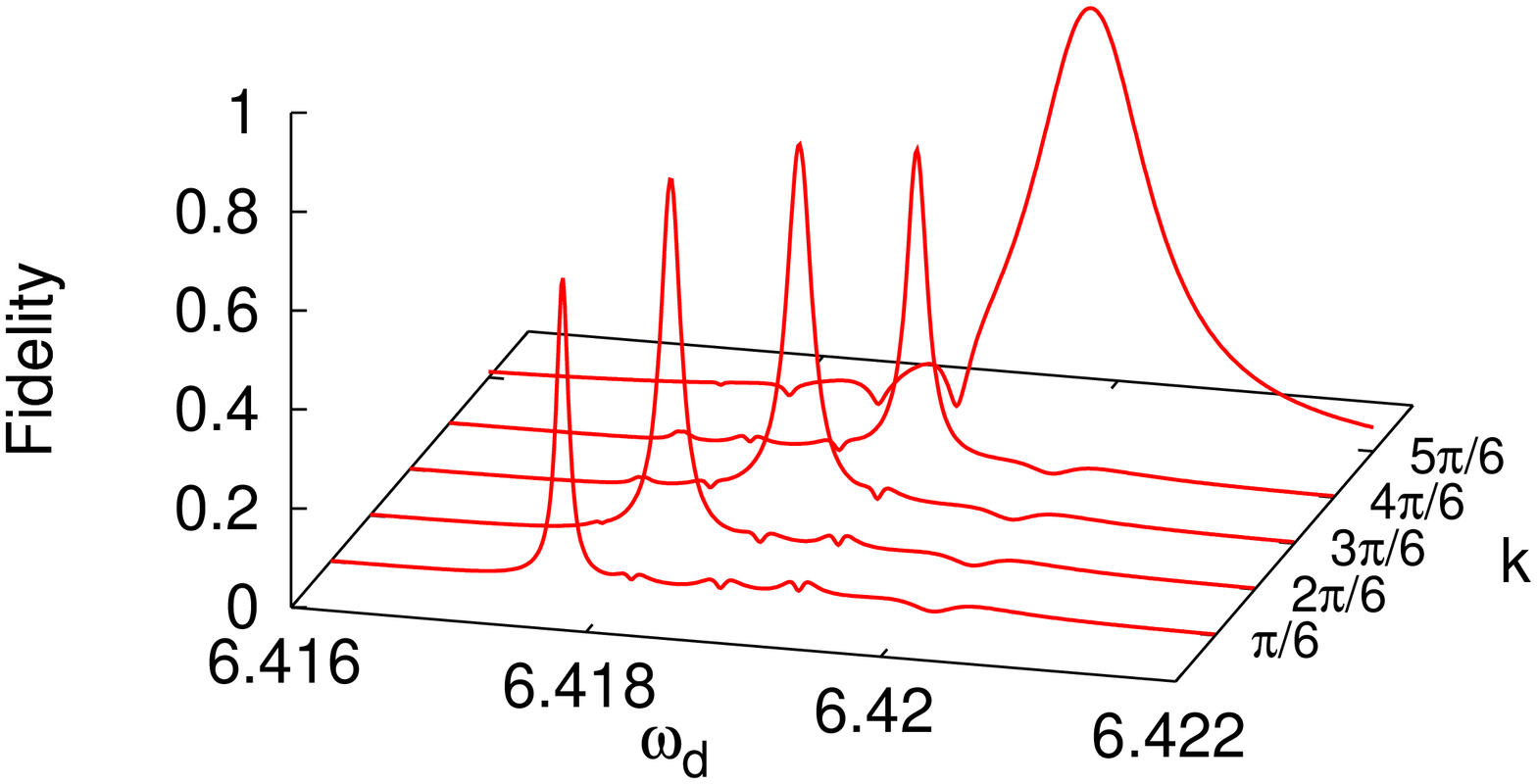} 
\vspace{-3cm}

\includegraphics[width=6.5cm, angle=-90]{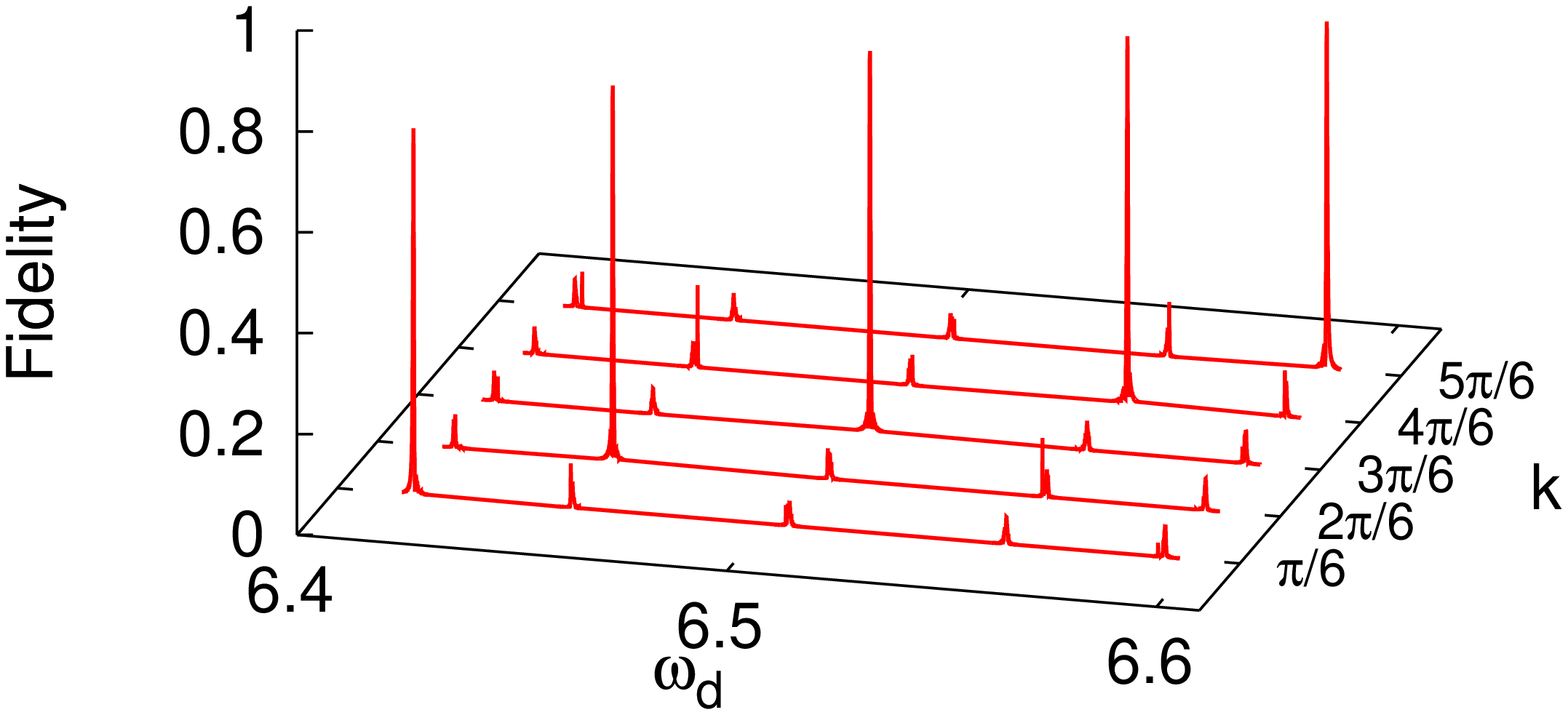} 
\caption{\footnotesize  \label{fig:firstsiteopen} (color online) Fidelities to all the possible $|k \rangle$ states versus drive frequency $\omega_\rmd$ for open boundary conditions. (Top panel) Only the first of the $N=5$ cavities is driven, the energy difference between driving to two neighboring $|k\rangle$ states is controlled by $\Delta \widetilde{E} \sim 2\pi J (g/\Delta)^2 /N$. (Bottom panel) However, when all cavities are driven equally, the energy difference is controlled by $\Delta \omega \sim 2\pi J  /N$, allowing a much better control over stabilizing the desired target state. Please note the widely different scales for the frequency ranges shown in the two plots.  
$\epsilon^\rmd = 0.3, \omega_\rmc = 6, \Delta = 1, g = J = 10^{-1}, \kappa=10^{-4}, \gamma=10^{-5}, \gamma_\phi = 10^{-6}$ in units of $2\pi$~GHz.}
\end{figure}

\paragraph*{Effective master equation simulations.}
To complement the analytic approach, we have performed full numerical simulations of the effective master equation~(\ref{eq:master}) where we (i) compute exactly the full spectrum of the spin chain $H_\sigma$ in Eq.~(\ref{eq:Hspinchain}) with $N=5$ qubits and open boundary conditions (ii) determine the rates between all the eigenstates and (iii) solve for the steady-state populations. In these calculations, sufficient number of higher-excitation manifolds of the spin chain were included to achieve convergence.

Because the parameter space is fairly large, we performed our simulations for a presently existing fabricated system for $N=2$~\cite{schwartz_toward_2015}. These parameters are quoted in the caption of Fig.~\ref{fig:firstsiteopen}, where the fidelities to achieve various spin-chain states ($k$) for $N=5$ are compared for a localized drive [Fig.~\ref{fig:firstsiteopen}(a)] and a spatially uniform drive [Fig.~\ref{fig:firstsiteopen}(b)]. We note that compared to the current state of the art~\cite{shankar_autonomously_2013}, these are remarkable fidelities. These fidelities can be significantly improved by reducing the ratio of the dephasing over the qubit relaxation rate. 

We have also tested the robustness of our protocol against site-to-site inhomogeneities of the different parameters and found no qualitative difference for $\delta\omega_\rmc/\omega_\rmc \sim 10^{-2}$, $\delta\omega_\rmq/\omega_\rmq \sim 10^{-4}$, $\delta g/g \sim 10^{-4}$ and $\delta J/J \sim 10^{-2}$. A more extensive analysis of achievable fidelities in the presence of site-to-site inhomogeneities will be presented in future work. 

\begin{figure}[t!]
\includegraphics[width=6.5cm, angle=-90]{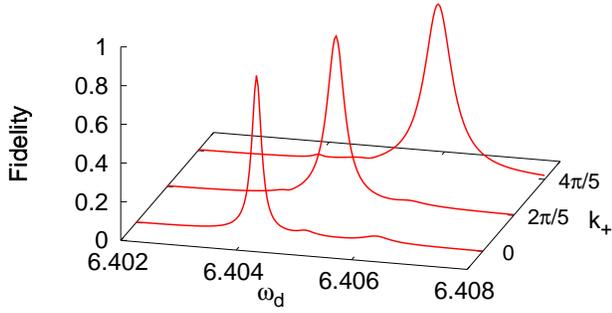} 
\caption{\footnotesize \label{fig:firstsite} (color online) Fidelities to the $|k_+\rangle$ states defined in Eq.~(\ref{eq:kpm}) versus drive frequency $\omega_\rmd$ in case only the first of $N=5$ cavities is driven and with periodic boundary conditions. Same parameters as in Fig.~\ref{fig:firstsiteopen}.
}
\end{figure}

\paragraph*{Periodic boundary conditions.}
Let us now consider the case of periodic boundary conditions for which the {\it undriven} system is space-translational and space-reversal invariant.
Such a symmetry results in degeneracies between the eigenstates $|k\rangle$ and $|2\pi -k \rangle$ of the undriven spin chain (except for $k=0$ and $k=\pi$). 
A symmetry-breaking drive profile will generically lift the degeneracy of the spectrum and, importantly, the emergent eigenstates will strongly depend on the particular drive profile.
To exemplify this point, let us start by driving the first cavity only: $\epsilon^\rmd_i = \epsilon^\rmd \delta_{i,0}$.
Second-order degenerate perturbation theory in $({g}/{\Delta})(\epsilon^\rmd/\sqrt{N}\Delta_\rmq)$ lifts the degeneracy in the subspaces spanned by  $|k\rangle$ and $|2\pi - k \rangle$. To lowest order, the eigenstates are
\begin{align} \label{eq:kpm}
|k_\pm \rangle \equiv \frac{| k \rangle \pm  | 2\pi - k \rangle}{\sqrt{2}}
\end{align}
for all $k \in\, ]0,\pi[$ complemented with the W state $|0_+\rangle \equiv | 0 \rangle$ and $|\pi_+\rangle \equiv | \pi \rangle$ (for $N$ even), one obtains the rates 
$\Gamma_{\boldsymbol{0} \to k_\pm} = 2 \pi   \Lambda_{k_\pm}^2 \sum_q \rho_{q}(\omega_\rmd+\widetilde{E}_{\boldsymbol{0}} -\widetilde{E}_{k_\pm})$
with 
\begin{align}
 \Lambda_{k_-}  \!= 0 \mbox{ and }
 \Lambda_{k_+} \! =  \!\left| \frac{\sqrt{2}}{N^2}  \left(1+2\frac{\Delta}{\Delta_\rmc} \right) 
\left(\frac{g}{\Delta}\right)^3  \frac{{(\epsilon^\rmd})^2}{\Delta_\rmq} \right|
 \end{align}
for all $|k_-\rangle$ and $|k_+\rangle$ except for $|0\rangle$ or $|\pi\rangle$ in which case $\Lambda_{k_+}$ is reduced by a factor $\sqrt{2}$. Figure~\ref{fig:firstsite} shows that the $|k_+\rangle$ states can be obtained with substantial fidelities.

It is worth noting that, in the case of a generic driving profile $\epsilon^\rmd_i$, instead of Eq.~(\ref{eq:kpm}), the emergent eigenstates are given by
\begin{align}
|k_\pm \rangle \equiv \frac{| k \rangle \pm  \alpha_k | 2\pi - k \rangle}{\sqrt{1+\alpha_k^2}}
\end{align}
where $\alpha_k$,  the relative weight of $|k\rangle$ and $|2\pi-k\rangle$, is now  controlled by the ratio $\epsilon^\rmd_k/\epsilon^\rmd_{2\pi-k}$. Therefore, such a non-equilibrium symmetry-breaking scenario offers highly flexible control over the target entangled state by simply engineering the drive profile $\epsilon^\rmd_k$.

\section{Photon-mediated interactions: General formulation}

\begin{figure}[!t]
\includegraphics[width=7.5cm]{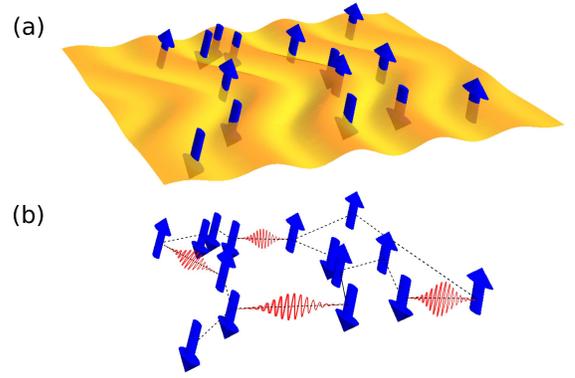} 
\caption{\label{fig:3dEM} \footnotesize (color online) Light-mediated interactions offer a highly versatile platform to design and control networks of interacting qubits. (a) The qubits (blue arrows) are embedded in an electromagnetic environment, which is the solution of the Maxwell equations in a given scattering geometry. (b) Integrating out the EM degrees of freedom yields effective interactions between the qubits, forming a network that can sustain large-scale entangled many-body states.} 
\end{figure}

\label{sec:ph-mediated}
In the previous sections, we focused on a particular geometry of the cavity-qubit system, namely a one-dimensional tight-binding lattice of photons. In this section, we show that our dissipative stabilization scheme via photon-mediated interactions is broadly applicable to any engineered EM environment. The situation we consider here is depicted in Fig.~6 where the two-level systems are now placed at arbitrary locations and interact with a general EM environment [Fig.~6(a)]. In practice, the latter is described by the solution of Maxwell equations in an arbitrary scattering geometry, characterized by a spectral problem with certain continuity and boundary conditions. In the simplest case, this would be a single resonator or a waveguide or, as in the specific example discussed previously, an array of evanescently (or capacitively) coupled cavities  (transmission line cavities). After the derivation of the most general result, we show how the effective tight-binding result can be derived from first principles. 

The qubits and their coupling to the EM fluctuations are described by the Hamiltonian 
\begin{align}
H =  H_{\sigma}  + H_{\sigma-{\rm EM}} + H_{\rm EM}\,,
\end{align}
with  ($\hbar = 1$)
\begin{align}
& H_{\sigma} = \sum_i \omega^\rmq_i \frac{\sigma^z_i}{2}, \quad 
H_{\rm EM} = \frac{\epsilon_0}{2} \int_\mathcal{V} \!\! \rmd^3 \boldsymbol{x} \, \left( \boldsymbol{E}^2 + c^2 \boldsymbol{B}^2 \right), \\
& H_{\sigma-{\rm EM}} =  - \int_\mathcal{V} \!\! \rmd^3 \boldsymbol{x} \,\boldsymbol{P} \cdot \boldsymbol{E}\,.
\end{align}
The integrals above run over the entire volume $\mathcal{V}$ of the scattering structure. We note that such a Hamiltonian can indeed be obtained for sub-gap electrodynamics in a general superconducting circuit architecture by a proper choice of normal modes~\cite{malekakhlagh_origin_2015}, starting from the parameters (position-dependent capacitances and inductances per unit length and the parameters of the qubits) of the underlying electrical circuit. 

For qubits residing at $\boldsymbol{x}_i$, each with a dipole moment strength $\mu$ (projected along a particular eigenpolarization of the electromagnetic medium), the collective atomic polarization operator can be written as
\begin{align}
P(\boldsymbol{x}) = \mu \sum_i  \delta^3(\boldsymbol{x}-\boldsymbol{x}_i) \, \sigma_i^x\,.
\end{align}
The electric field can generally be written in terms of a complete set of modes and corresponding eigenfrequencies $\{ \varphi_n , \omega_n \}$ specific to the chosen architecture
\begin{align}
E(\boldsymbol{x}) = \sum_{n}  \varepsilon_n \varphi_n(\boldsymbol{x}) a_{n} + \mbox{H.c.}\;,
\end{align}
so that
\begin{align}\label{eq:Ha1}
H_{\rm EM} = \sum_{n} \omega_n a_{n}^\dagger  a_{n}\,.
\end{align}
Here, $a^\dagger_n$ ($a_n$) creates (annihilates) a photon in the spatial mode $\varphi_n(\boldsymbol{x})$ with frequency $\omega_n$ and corresponding zero-point electric field $\varepsilon_n$. The modes are assumed to satisfy the completeness and orthogonality conditions $\sum_n \varphi_n(\boldsymbol{x}) \varphi_n ^* (\boldsymbol{x}') = \delta^3(\boldsymbol{x}-\boldsymbol{x}')$ and $ \int_\mathcal{V} \! \rmd^3 \boldsymbol{x} \, \varphi_n(\boldsymbol{x}) \varphi_m^*(\boldsymbol{x}) = \delta_{nm} $. With these normalization conditions, the zero-point fields are given by $\varepsilon_n \equiv \sqrt{{\omega_n}/{2\epsilon_0}}$.
Neglecting counter-rotating terms, the light-matter coupling becomes
\begin{align}
H_{\sigma-{\rm EM}} = &  \sum_{i,n} g_n \varphi_n(\boldsymbol{x}_i)\,  a_n^\dagger \sigma_i^- + \mbox{H.c.}\,,
\end{align}
where $\sigma_i^\pm \equiv (\sigma^x_i \pm \rmi\sigma_i^y)/2$ and $g_n \equiv - \mu \varepsilon_n$.

We shall consider the regime in which the qubit frequencies $\omega^\rmq_i$ are far detuned from the photonic modes $\omega_n$ such that the light-matter coupling can be treated via second-order perturbation theory in $g_n/(\omega^\rmq_i - \omega_n)$. This is achieved by means of a Schrieffer-Wolf (SW) transformation~\cite{schrieffer_relation_1966} which maps $H \mapsto \rme^{X} H  \rme^{X^\dagger}$ where
\begin{align} \label{eq:SWX}
X \equiv  \sum_{n,i}  \left[ \frac{g_n \varphi_n(\boldsymbol{x}_i)}{\omega_i^\rmq - \omega_n} \sigma_i^+ a_n 
-\mathrm{H.c.}
 \right]. 
\end{align}
This yields the following Hamiltonian $H = H_\sigma + H_{\sigma a} + H_a$ to $\mathcal{O}(g^2/\Delta^2)$, with
\begin{align}
H_{\sigma} =& \sum_i \omega^\rmq_i \frac{\sigma^z_i}{2} + \frac{1}{2} \sum_{ij} \Sigma_{ij}(\omega^\rmq_i)  \sigma_j^- \sigma_i^+ + \mbox{H.c.}\,, \label{eq:hsigma} \\
H_{\sigma a} =& \sum_{i, mn}  \lambda_{i, mn}(\omega_i^\rmq)  \,\frac{\sigma^z_i}{2} \, a_n a^\dagger_m   + \mbox{H.c.} \,, \label{eq:sigmaa}
\end{align}
and $H_a$ is still given by Eq.~(\ref{eq:Ha1}). In this low-energy Hamiltonian, the second term in $H_\sigma$ describes the qubit-qubit interactions mediated by virtual photons. These photons can be emitted into and absorbed from photonic channels at frequencies $\omega_n$ with spatial distribution $\varphi_n(\boldsymbol{x})$. This is precisely the story told by the coefficients $\Sigma_{ij}(\omega) =  \sum_n |g_n|^2 \varphi_n(\boldsymbol{x}_i) \varphi_n^* (\boldsymbol{x}_j)/(\omega - \omega_n)$, which appears as a self-energy correction to the qubit sector. $\Sigma_{ij}(\omega) \sigma_j^x$ can be seen as 
the electric field generated at $\boldsymbol{x}_i$ by a dipole at $\boldsymbol{x}_j$, oscillating harmonically at frequency $\omega$. Note that the bare electromagnetic retarded Green's function is given by $G^{\rm R} (\boldsymbol{x},\boldsymbol{x}'; \omega) = \sum_n \varphi_n(\boldsymbol{x}) \varphi_n^*(\boldsymbol{x}')/(\omega - \omega_n)$. This immediately implies that, in principle, all qubits interact with each other, to the extent that they can radiate EM radiation to each other [see schematic in Fig.~6(b)]. We note that realization-specific and restricted versions of this interaction vertex have been derived before~\cite{majer_coupling_2007,loo_photon-mediated_2013}. 

For what is proposed here, however, an equally important role is played by the term $H_{\sigma a}$ in Eq.~(\ref{eq:sigmaa}). This is the generalized version of the ac Stark-shift contribution to a qubit's frequency that is well known in the dispersive regime of single-mode Cavity QED~\cite{boissonneault_dispersive_2009}, which can also be interpreted as a scattering term for photons generated by the interaction of the radiation field with qubits. This term expresses the spatial fluctuations of the effective index of refraction of the electromagnetic medium through the dynamically generated polarization fluctuations [\textit{i.e.} $P(\boldsymbol{x})$] of the qubits. The interaction vertex here is again given by the resonant modes and their frequencies: $\lambda_{i,mn}(\omega) = g_n g_m^* \varphi_n (\boldsymbol{x}_i) \varphi_m^*(\boldsymbol{x}_i)/(\omega - \omega_n)$. We note that for a 1D tight-binding model of a cavity array with nearest-neighbor hopping, $\Sigma_{ij} (\omega_\rmq) \simeq  (g/\Delta)^2 [ \Delta \delta_{i,j} - J\delta_{i,j\pm 1}] $ and $\lambda_{i,kq} (\omega_\rmq) \simeq \Delta (g/\Delta)^2 \varphi_k^*(i)  \varphi_q(i)$ to lowest relevant order in $J/\Delta$ and $g/\Delta$. This result agrees with our direct derivation in Sec.~\ref{sec:cooling}. These results can be extended to a full-fledged stabilization protocol for a generalized W state $| W_n \rangle =  \sum_{i=0}^{N-1}  \varphi_n^* (\boldsymbol{x}_i) |\downarrow_0 \ldots\downarrow_{i-1} \, \uparrow_i \, \downarrow_{i+1} \ldots \downarrow_{N-1}  \rangle$.

\section{Discussion and Conclusions}
\label{sec:end}

We proposed a general and scalable method based on photon-mediated interactions to drive a set of $N$ qubits to a desired generalized W state. The particular protocol discussed here for qubits embedded in a cavity array, amounts to the dissipative stabilization of a particular {\it excited state} of a many-body system (in the present case, a non-integrable variant of the XY model). This approach stands in contrast to cooling techniques employed for condensed-matter and cold atomic systems that, at least in principle, target the stabilization of the ground state. 

An interesting feature of the present approach compared to earlier approaches to dissipative engineering of entanglement~\cite{kraus_preparation_2008, diehl_quantum_2008, cormick_dissipative_2013,lee_emergence_2013, rao_deterministic_2014,reiter_scalable_2015} is the fact that {\it both} the unitary and the dissipative parts of the dynamics are adjusted through coupling to a common photonic bath. The Hamiltonian part provides the set of pure many-body states that can be reached in the steady state, while the dissipative part determines the occupation of those states. For cavities that are high-Q, the transitions can be made very selective. In the case of the dissipative stabilization of a W state of $N$ qubits, we discussed the scaling of the fidelity with the system size $N$.  

The fact that various properties of multi-qubit dynamics can be precisely adjusted by drive parameters provides a suitable platform for quantum simulation~\cite{rotondo_dicke_2015} and computation~\cite{verstraete_quantum_2009}. In scaling up CQED-based simulators~{\cite{PhysRevLett.115.240501}} to larger architectures, one of the main obstacles is the uncontrolled site-to-site fluctuation of system parameters~\cite{underwood_low-disorder_2012}. In the presented scheme, the {\it dynamical tuning} of effective spin-chain parameters, in fact both the unitary and the dissipative parameters, through the drive frequency and amplitude provides a promising route to realize large-scale quantum simulators.  

More generally, the method proposed here and its possible generalization to higher dimensional lattices holds promise for various quantum information applications, such as deterministic teleportation \cite{agrawal_perfect_2006, wang_simple_2009}. The reduction of the collective dephasing mechanisms and the generation of entangled states in higher-excitation manifolds are important goals. Another interesting open question is the adaptation and extension of our protocol for targeted many-body state preparation in the photonic sector, scaling up recent approaches~\cite{leghtas_confining_2015}.

\section{Acknowledgements}
\label{sec:ack}

We are grateful to Mollie Schwartz, Leigh Martin, Emmanuel Flurin, Irfan Siddiqi, Liang Jiang, and Alexandre Blais for helpful discussions. This work has been supported by ARO Grant No. W911NF-15-1-0299 and NSF Grant No. DMR-1151810. M.K gratefully acknowledges support from the Professional Staff Congress of the City University of New York award No. 68193-0046.

\bibliography{SpinChain}

\begin{thebibliography}{51}%
\makeatletter
\providecommand \@ifxundefined [1]{%
 \@ifx{#1\undefined}
}%
\providecommand \@ifnum [1]{%
 \ifnum #1\expandafter \@firstoftwo
 \else \expandafter \@secondoftwo
 \fi
}%
\providecommand \@ifx [1]{%
 \ifx #1\expandafter \@firstoftwo
 \else \expandafter \@secondoftwo
 \fi
}%
\providecommand \natexlab [1]{#1}%
\providecommand \enquote  [1]{``#1''}%
\providecommand \bibnamefont  [1]{#1}%
\providecommand \bibfnamefont [1]{#1}%
\providecommand \citenamefont [1]{#1}%
\providecommand \href@noop [0]{\@secondoftwo}%
\providecommand \href [0]{\begingroup \@sanitize@url \@href}%
\providecommand \@href[1]{\@@startlink{#1}\@@href}%
\providecommand \@@href[1]{\endgroup#1\@@endlink}%
\providecommand \@sanitize@url [0]{\catcode `\\12\catcode `\$12\catcode
  `\&12\catcode `\#12\catcode `\^12\catcode `\_12\catcode `\%12\relax}%
\providecommand \@@startlink[1]{}%
\providecommand \@@endlink[0]{}%
\providecommand \url  [0]{\begingroup\@sanitize@url \@url }%
\providecommand \@url [1]{\endgroup\@href {#1}{\urlprefix }}%
\providecommand \urlprefix  [0]{URL }%
\providecommand \Eprint [0]{\href }%
\providecommand \doibase [0]{http://dx.doi.org/}%
\providecommand \selectlanguage [0]{\@gobble}%
\providecommand \bibinfo  [0]{\@secondoftwo}%
\providecommand \bibfield  [0]{\@secondoftwo}%
\providecommand \translation [1]{[#1]}%
\providecommand \BibitemOpen [0]{}%
\providecommand \bibitemStop [0]{}%
\providecommand \bibitemNoStop [0]{.\EOS\space}%
\providecommand \EOS [0]{\spacefactor3000\relax}%
\providecommand \BibitemShut  [1]{\csname bibitem#1\endcsname}%
\let\auto@bib@innerbib\@empty
\bibitem [{\citenamefont {Feynman}(1950)}]{feynman_mathematical_1950}%
  \BibitemOpen
  \bibfield  {author} {\bibinfo {author} {\bibfnamefont {R.~P.}\ \bibnamefont
  {Feynman}},\ }\bibfield  {title} {\enquote {\bibinfo {title} {{Mathematical
  {Formulation} of the {Quantum} {Theory} of {Electromagnetic}
  {Interaction}}},}\ }\href {\doibase 10.1103/PhysRev.80.440} {\bibfield
  {journal} {\bibinfo  {journal} {Phys. Rev.}\ }\textbf {\bibinfo {volume}
  {80}},\ \bibinfo {pages} {440--457} (\bibinfo {year} {1950})}\BibitemShut
  {NoStop}%
\bibitem [{\citenamefont {Schwinger}(1975)}]{schwinger_casimir_1975}%
  \BibitemOpen
  \bibfield  {author} {\bibinfo {author} {\bibfnamefont {Julian}\ \bibnamefont
  {Schwinger}},\ }\bibfield  {title} {{\selectlanguage {english}\enquote
  {\bibinfo {title} {{Casimir effect in source theory}},}\ }}\href {\doibase
  10.1007/BF00405585} {\bibfield  {journal} {\bibinfo  {journal} {Lett Math
  Phys}\ }\textbf {\bibinfo {volume} {1}},\ \bibinfo {pages} {43--47} (\bibinfo
  {year} {1975})}\BibitemShut {NoStop}%
\bibitem [{\citenamefont {Dicke}(1954)}]{dicke_coherence_1954}%
  \BibitemOpen
  \bibfield  {author} {\bibinfo {author} {\bibfnamefont {R.~H.}\ \bibnamefont
  {Dicke}},\ }\bibfield  {title} {\enquote {\bibinfo {title} {{Coherence in
  {Spontaneous} {Radiation} {Processes}}},}\ }\href {\doibase
  10.1103/PhysRev.93.99} {\bibfield  {journal} {\bibinfo  {journal} {Phys.
  Rev.}\ }\textbf {\bibinfo {volume} {93}},\ \bibinfo {pages} {99--110}
  (\bibinfo {year} {1954})}\BibitemShut {NoStop}%
\bibitem [{\citenamefont {Gross}\ and\ \citenamefont
  {Haroche}(1982)}]{gross_superradiance_1982}%
  \BibitemOpen
  \bibfield  {author} {\bibinfo {author} {\bibfnamefont {M.}~\bibnamefont
  {Gross}}\ and\ \bibinfo {author} {\bibfnamefont {S.}~\bibnamefont
  {Haroche}},\ }\bibfield  {title} {\enquote {\bibinfo {title} {Superradiance:
  An essay on the theory of collective spontaneous emission},}\ }\href
  {\doibase 10.1016/0370-1573(82)90102-8} {\bibfield  {journal} {\bibinfo
  {journal} {Phys. Rep.}\ }\textbf {\bibinfo {volume} {93}},\ \bibinfo {pages}
  {301--396} (\bibinfo {year} {1982})}\BibitemShut {NoStop}%
\bibitem [{\citenamefont {Brandes}(2005)}]{brandes_coherent_2005}%
  \BibitemOpen
  \bibfield  {author} {\bibinfo {author} {\bibfnamefont {Tobias}\ \bibnamefont
  {Brandes}},\ }\bibfield  {title} {\enquote {\bibinfo {title} {Coherent and
  collective quantum optical effects in mesoscopic systems},}\ }\href {\doibase
  10.1016/j.physrep.2004.12.002} {\bibfield  {journal} {\bibinfo  {journal}
  {Phys. Rep.}\ }\textbf {\bibinfo {volume} {408}},\ \bibinfo {pages}
  {315--474} (\bibinfo {year} {2005})}\BibitemShut {NoStop}%
\bibitem [{\citenamefont {DeVoe}\ and\ \citenamefont
  {Brewer}(1996)}]{devoe_observation_1996}%
  \BibitemOpen
  \bibfield  {author} {\bibinfo {author} {\bibfnamefont {R.~G.}\ \bibnamefont
  {DeVoe}}\ and\ \bibinfo {author} {\bibfnamefont {R.~G.}\ \bibnamefont
  {Brewer}},\ }\bibfield  {title} {\enquote {\bibinfo {title} {{Observation of
  {Superradiant} and {Subradiant} {Spontaneous} {Emission} of {Two} {Trapped}
  {Ions}}},}\ }\href {\doibase 10.1103/PhysRevLett.76.2049} {\bibfield
  {journal} {\bibinfo  {journal} {Phys. Rev. Lett.}\ }\textbf {\bibinfo
  {volume} {76}},\ \bibinfo {pages} {2049--2052} (\bibinfo {year}
  {1996})}\BibitemShut {NoStop}%
\bibitem [{\citenamefont {Eschner}\ \emph {et~al.}(2001)\citenamefont
  {Eschner}, \citenamefont {Raab}, \citenamefont {Schmidt-Kaler},\ and\
  \citenamefont {Blatt}}]{eschner_light_2001}%
  \BibitemOpen
  \bibfield  {author} {\bibinfo {author} {\bibfnamefont {J.}~\bibnamefont
  {Eschner}}, \bibinfo {author} {\bibfnamefont {Ch}~\bibnamefont {Raab}},
  \bibinfo {author} {\bibfnamefont {F.}~\bibnamefont {Schmidt-Kaler}}, \ and\
  \bibinfo {author} {\bibfnamefont {R.}~\bibnamefont {Blatt}},\ }\bibfield
  {title} {{\selectlanguage {english}\enquote {\bibinfo {title} {{Light
  interference from single atoms and their mirror images}},}\ }}\href {\doibase
  10.1038/35097017} {\bibfield  {journal} {\bibinfo  {journal} {Nature}\
  }\textbf {\bibinfo {volume} {413}},\ \bibinfo {pages} {495--498} (\bibinfo
  {year} {2001})}\BibitemShut {NoStop}%
\bibitem [{\citenamefont {Ritsch}\ \emph {et~al.}(2013)\citenamefont {Ritsch},
  \citenamefont {Domokos}, \citenamefont {Brennecke},\ and\ \citenamefont
  {Esslinger}}]{ritsch_cold_2013}%
  \BibitemOpen
  \bibfield  {author} {\bibinfo {author} {\bibfnamefont {Helmut}\ \bibnamefont
  {Ritsch}}, \bibinfo {author} {\bibfnamefont {Peter}\ \bibnamefont {Domokos}},
  \bibinfo {author} {\bibfnamefont {Ferdinand}\ \bibnamefont {Brennecke}}, \
  and\ \bibinfo {author} {\bibfnamefont {Tilman}\ \bibnamefont {Esslinger}},\
  }\bibfield  {title} {\enquote {\bibinfo {title} {{Cold atoms in
  cavity-generated dynamical optical potentials}},}\ }\href {\doibase
  10.1103/RevModPhys.85.553} {\bibfield  {journal} {\bibinfo  {journal}
  {Reviews of Modern Physics}\ }\textbf {\bibinfo {volume} {85}},\ \bibinfo
  {pages} {553--601} (\bibinfo {year} {2013})}\BibitemShut {NoStop}%
\bibitem [{\citenamefont {Baumann}\ \emph {et~al.}(2010)\citenamefont
  {Baumann}, \citenamefont {Guerlin}, \citenamefont {Brennecke},\ and\
  \citenamefont {Esslinger}}]{baumann_dicke_2010}%
  \BibitemOpen
  \bibfield  {author} {\bibinfo {author} {\bibfnamefont {Kristian}\
  \bibnamefont {Baumann}}, \bibinfo {author} {\bibfnamefont {Christine}\
  \bibnamefont {Guerlin}}, \bibinfo {author} {\bibfnamefont {Ferdinand}\
  \bibnamefont {Brennecke}}, \ and\ \bibinfo {author} {\bibfnamefont {Tilman}\
  \bibnamefont {Esslinger}},\ }\bibfield  {title} {{\selectlanguage
  {english}\enquote {\bibinfo {title} {{Dicke quantum phase transition with a
  superfluid gas in an optical cavity}},}\ }}\href {\doibase
  10.1038/nature09009} {\bibfield  {journal} {\bibinfo  {journal} {Nature}\
  }\textbf {\bibinfo {volume} {464}},\ \bibinfo {pages} {1301--1306} (\bibinfo
  {year} {2010})}\BibitemShut {NoStop}%
\bibitem [{\citenamefont {Mottl}\ \emph {et~al.}(2012)\citenamefont {Mottl},
  \citenamefont {Brennecke}, \citenamefont {Baumann}, \citenamefont {Landig},
  \citenamefont {Donner},\ and\ \citenamefont
  {Esslinger}}]{mottl_roton-type_2012}%
  \BibitemOpen
  \bibfield  {author} {\bibinfo {author} {\bibfnamefont {R.}~\bibnamefont
  {Mottl}}, \bibinfo {author} {\bibfnamefont {F.}~\bibnamefont {Brennecke}},
  \bibinfo {author} {\bibfnamefont {K.}~\bibnamefont {Baumann}}, \bibinfo
  {author} {\bibfnamefont {R.}~\bibnamefont {Landig}}, \bibinfo {author}
  {\bibfnamefont {T.}~\bibnamefont {Donner}}, \ and\ \bibinfo {author}
  {\bibfnamefont {T.}~\bibnamefont {Esslinger}},\ }\bibfield  {title}
  {{\selectlanguage {english}\enquote {\bibinfo {title} {{Roton-{Type} {Mode}
  {Softening} in a {Quantum} {Gas} with {Cavity}-{Mediated} {Long}-{Range}
  {Interactions}}},}\ }}\href {\doibase 10.1126/science.1220314} {\bibfield
  {journal} {\bibinfo  {journal} {Science}\ }\textbf {\bibinfo {volume}
  {336}},\ \bibinfo {pages} {1570--1573} (\bibinfo {year} {2012})}\BibitemShut
  {NoStop}%
\bibitem [{\citenamefont {Brennecke}\ \emph {et~al.}(2013)\citenamefont
  {Brennecke}, \citenamefont {Mottl}, \citenamefont {Baumann}, \citenamefont
  {Landig}, \citenamefont {Donner},\ and\ \citenamefont
  {Esslinger}}]{brennecke_real-time_2013}%
  \BibitemOpen
  \bibfield  {author} {\bibinfo {author} {\bibfnamefont {Ferdinand}\
  \bibnamefont {Brennecke}}, \bibinfo {author} {\bibfnamefont {Rafael}\
  \bibnamefont {Mottl}}, \bibinfo {author} {\bibfnamefont {Kristian}\
  \bibnamefont {Baumann}}, \bibinfo {author} {\bibfnamefont {Renate}\
  \bibnamefont {Landig}}, \bibinfo {author} {\bibfnamefont {Tobias}\
  \bibnamefont {Donner}}, \ and\ \bibinfo {author} {\bibfnamefont {Tilman}\
  \bibnamefont {Esslinger}},\ }\bibfield  {title} {{\selectlanguage
  {english}\enquote {\bibinfo {title} {{Real-time observation of fluctuations
  at the driven-dissipative {Dicke} phase transition}},}\ }}\href {\doibase
  10.1073/pnas.1306993110} {\bibfield  {journal} {\bibinfo  {journal}
  {Proceedings of the National Academy of Sciences}\ }\textbf {\bibinfo
  {volume} {110}},\ \bibinfo {pages} {11763--11767} (\bibinfo {year}
  {2013})}\BibitemShut {NoStop}%
\bibitem [{\citenamefont {Strack}\ and\ \citenamefont
  {Sachdev}(2011)}]{strack_dicke_2011}%
  \BibitemOpen
  \bibfield  {author} {\bibinfo {author} {\bibfnamefont {Philipp}\ \bibnamefont
  {Strack}}\ and\ \bibinfo {author} {\bibfnamefont {Subir}\ \bibnamefont
  {Sachdev}},\ }\bibfield  {title} {\enquote {\bibinfo {title} {{Dicke
  {Quantum} {Spin} {Glass} of {Atoms} and {Photons}}},}\ }\href {\doibase
  10.1103/PhysRevLett.107.277202} {\bibfield  {journal} {\bibinfo  {journal}
  {Physical Review Letters}\ }\textbf {\bibinfo {volume} {107}},\ \bibinfo
  {pages} {277202} (\bibinfo {year} {2011})}\BibitemShut {NoStop}%
\bibitem [{\citenamefont {Kulkarni}\ \emph {et~al.}(2013)\citenamefont
  {Kulkarni}, \citenamefont {{\"O}ztop},\ and\ \citenamefont
  {T{\"u}reci}}]{kulkarni_cavity-mediated_2013}%
  \BibitemOpen
  \bibfield  {author} {\bibinfo {author} {\bibfnamefont {Manas}\ \bibnamefont
  {Kulkarni}}, \bibinfo {author} {\bibfnamefont {Baris}\ \bibnamefont
  {{\"O}ztop}}, \ and\ \bibinfo {author} {\bibfnamefont {Hakan~E.}\
  \bibnamefont {T{\"u}reci}},\ }\bibfield  {title} {\enquote {\bibinfo {title}
  {{Cavity-{Mediated} {Near}-{Critical} {Dissipative} {Dynamics} of a {Driven}
  {Condensate}}},}\ }\href {\doibase 10.1103/PhysRevLett.111.220408} {\bibfield
   {journal} {\bibinfo  {journal} {Phys. Rev. Lett.}\ }\textbf {\bibinfo
  {volume} {111}},\ \bibinfo {pages} {220408} (\bibinfo {year}
  {2013})}\BibitemShut {NoStop}%
\bibitem [{\citenamefont {Konya}\ \emph {et~al.}(2014)\citenamefont {Konya},
  \citenamefont {Szirmai},\ and\ \citenamefont {Domokos}}]{konya_damping_2014}%
  \BibitemOpen
  \bibfield  {author} {\bibinfo {author} {\bibfnamefont {G.}~\bibnamefont
  {Konya}}, \bibinfo {author} {\bibfnamefont {G.}~\bibnamefont {Szirmai}}, \
  and\ \bibinfo {author} {\bibfnamefont {P.}~\bibnamefont {Domokos}},\
  }\bibfield  {title} {\enquote {\bibinfo {title} {{Damping of quasiparticles
  in a {Bose}-{Einstein} condensate coupled to an optical cavity}},}\ }\href
  {\doibase 10.1103/PhysRevA.90.013623} {\bibfield  {journal} {\bibinfo
  {journal} {Physical Review A}\ }\textbf {\bibinfo {volume} {90}},\ \bibinfo
  {pages} {013623} (\bibinfo {year} {2014})}\BibitemShut {NoStop}%
\bibitem [{\citenamefont {Majer}\ \emph {et~al.}(2007)\citenamefont {Majer},
  \citenamefont {Chow}, \citenamefont {Gambetta}, \citenamefont {Koch},
  \citenamefont {Johnson}, \citenamefont {Schreier}, \citenamefont {Frunzio},
  \citenamefont {Schuster}, \citenamefont {Houck}, \citenamefont {Wallraff},
  \citenamefont {Blais}, \citenamefont {Devoret}, \citenamefont {Girvin},\ and\
  \citenamefont {Schoelkopf}}]{majer_coupling_2007}%
  \BibitemOpen
  \bibfield  {author} {\bibinfo {author} {\bibfnamefont {J.}~\bibnamefont
  {Majer}}, \bibinfo {author} {\bibfnamefont {J.~M.}\ \bibnamefont {Chow}},
  \bibinfo {author} {\bibfnamefont {J.~M.}\ \bibnamefont {Gambetta}}, \bibinfo
  {author} {\bibfnamefont {Jens}\ \bibnamefont {Koch}}, \bibinfo {author}
  {\bibfnamefont {B.~R.}\ \bibnamefont {Johnson}}, \bibinfo {author}
  {\bibfnamefont {J.~A.}\ \bibnamefont {Schreier}}, \bibinfo {author}
  {\bibfnamefont {L.}~\bibnamefont {Frunzio}}, \bibinfo {author} {\bibfnamefont
  {D.~I.}\ \bibnamefont {Schuster}}, \bibinfo {author} {\bibfnamefont {A.~A.}\
  \bibnamefont {Houck}}, \bibinfo {author} {\bibfnamefont {A.}~\bibnamefont
  {Wallraff}}, \bibinfo {author} {\bibfnamefont {A.}~\bibnamefont {Blais}},
  \bibinfo {author} {\bibfnamefont {M.~H.}\ \bibnamefont {Devoret}}, \bibinfo
  {author} {\bibfnamefont {S.~M.}\ \bibnamefont {Girvin}}, \ and\ \bibinfo
  {author} {\bibfnamefont {R.~J.}\ \bibnamefont {Schoelkopf}},\ }\bibfield
  {title} {{\selectlanguage {english}\enquote {\bibinfo {title} {{Coupling
  superconducting qubits via a cavity bus}},}\ }}\href {\doibase
  10.1038/nature06184} {\bibfield  {journal} {\bibinfo  {journal} {Nature}\
  }\textbf {\bibinfo {volume} {449}},\ \bibinfo {pages} {443--447} (\bibinfo
  {year} {2007})}\BibitemShut {NoStop}%
\bibitem [{\citenamefont {Filipp}\ \emph {et~al.}(2011)\citenamefont {Filipp},
  \citenamefont {G{\"o}ppl}, \citenamefont {Fink}, \citenamefont {Baur},
  \citenamefont {Bianchetti}, \citenamefont {Steffen},\ and\ \citenamefont
  {Wallraff}}]{filipp_multimode_2011}%
  \BibitemOpen
  \bibfield  {author} {\bibinfo {author} {\bibfnamefont {S.}~\bibnamefont
  {Filipp}}, \bibinfo {author} {\bibfnamefont {M.}~\bibnamefont {G{\"o}ppl}},
  \bibinfo {author} {\bibfnamefont {J.~M.}\ \bibnamefont {Fink}}, \bibinfo
  {author} {\bibfnamefont {M.}~\bibnamefont {Baur}}, \bibinfo {author}
  {\bibfnamefont {R.}~\bibnamefont {Bianchetti}}, \bibinfo {author}
  {\bibfnamefont {L.}~\bibnamefont {Steffen}}, \ and\ \bibinfo {author}
  {\bibfnamefont {A.}~\bibnamefont {Wallraff}},\ }\bibfield  {title} {\enquote
  {\bibinfo {title} {{Multimode mediated qubit-qubit coupling and dark-state
  symmetries in circuit quantum electrodynamics}},}\ }\href {\doibase
  10.1103/PhysRevA.83.063827} {\bibfield  {journal} {\bibinfo  {journal} {Phys.
  Rev. A}\ }\textbf {\bibinfo {volume} {83}},\ \bibinfo {pages} {063827}
  (\bibinfo {year} {2011})}\BibitemShut {NoStop}%
\bibitem [{\citenamefont {Loo}\ \emph {et~al.}(2013)\citenamefont {Loo},
  \citenamefont {Fedorov}, \citenamefont {Lalumi{\`e}re}, \citenamefont
  {Sanders}, \citenamefont {Blais},\ and\ \citenamefont
  {Wallraff}}]{loo_photon-mediated_2013}%
  \BibitemOpen
  \bibfield  {author} {\bibinfo {author} {\bibfnamefont {Arjan F.~van}\
  \bibnamefont {Loo}}, \bibinfo {author} {\bibfnamefont {Arkady}\ \bibnamefont
  {Fedorov}}, \bibinfo {author} {\bibfnamefont {Kevin}\ \bibnamefont
  {Lalumi{\`e}re}}, \bibinfo {author} {\bibfnamefont {Barry~C.}\ \bibnamefont
  {Sanders}}, \bibinfo {author} {\bibfnamefont {Alexandre}\ \bibnamefont
  {Blais}}, \ and\ \bibinfo {author} {\bibfnamefont {Andreas}\ \bibnamefont
  {Wallraff}},\ }\bibfield  {title} {{\selectlanguage {english}\enquote
  {\bibinfo {title} {{Photon-{Mediated} {Interactions} {Between} {Distant}
  {Artificial} {Atoms}}},}\ }}\href {\doibase 10.1126/science.1244324}
  {\bibfield  {journal} {\bibinfo  {journal} {Science}\ }\textbf {\bibinfo
  {volume} {342}},\ \bibinfo {pages} {1494--1496} (\bibinfo {year}
  {2013})}\BibitemShut {NoStop}%
\bibitem [{mly()}]{mlynek_observation_2014}%
  \BibitemOpen
  \bibfield  {title} {\enquote {\bibinfo {title} {Observation of {Dicke
  superradiance for two artificial atoms in a cavity with high decay rate},
  volume = {5}, journal = {Nat Commun}, author = {Mlynek, J. A. and
  Abdumalikov, A. A. and Eichler, C. and Wallraff, A.}, year = {2014}},}\
  }\href@noop {} {\ }\BibitemShut {NoStop}%
\bibitem [{\citenamefont {Murch}\ \emph {et~al.}(2012)\citenamefont {Murch},
  \citenamefont {Vool}, \citenamefont {Zhou}, \citenamefont {Weber},
  \citenamefont {Girvin},\ and\ \citenamefont
  {Siddiqi}}]{murch_cavity-assisted_2012}%
  \BibitemOpen
  \bibfield  {author} {\bibinfo {author} {\bibfnamefont {K.~W.}\ \bibnamefont
  {Murch}}, \bibinfo {author} {\bibfnamefont {U.}~\bibnamefont {Vool}},
  \bibinfo {author} {\bibfnamefont {D.}~\bibnamefont {Zhou}}, \bibinfo {author}
  {\bibfnamefont {S.~J.}\ \bibnamefont {Weber}}, \bibinfo {author}
  {\bibfnamefont {S.~M.}\ \bibnamefont {Girvin}}, \ and\ \bibinfo {author}
  {\bibfnamefont {I.}~\bibnamefont {Siddiqi}},\ }\bibfield  {title} {\enquote
  {\bibinfo {title} {{Cavity-{Assisted} {Quantum} {Bath} {Engineering}}},}\
  }\href {\doibase 10.1103/PhysRevLett.109.183602} {\bibfield  {journal}
  {\bibinfo  {journal} {Phys. Rev. Lett.}\ }\textbf {\bibinfo {volume} {109}},\
  \bibinfo {pages} {183602} (\bibinfo {year} {2012})}\BibitemShut {NoStop}%
\bibitem [{\citenamefont {Shankar}\ \emph {et~al.}(2013)\citenamefont
  {Shankar}, \citenamefont {Hatridge}, \citenamefont {Leghtas}, \citenamefont
  {Sliwa}, \citenamefont {Narla}, \citenamefont {Vool}, \citenamefont {Girvin},
  \citenamefont {Frunzio}, \citenamefont {Mirrahimi},\ and\ \citenamefont
  {Devoret}}]{shankar_autonomously_2013}%
  \BibitemOpen
  \bibfield  {author} {\bibinfo {author} {\bibfnamefont {S.}~\bibnamefont
  {Shankar}}, \bibinfo {author} {\bibfnamefont {M.}~\bibnamefont {Hatridge}},
  \bibinfo {author} {\bibfnamefont {Z.}~\bibnamefont {Leghtas}}, \bibinfo
  {author} {\bibfnamefont {K.~M.}\ \bibnamefont {Sliwa}}, \bibinfo {author}
  {\bibfnamefont {A.}~\bibnamefont {Narla}}, \bibinfo {author} {\bibfnamefont
  {U.}~\bibnamefont {Vool}}, \bibinfo {author} {\bibfnamefont {S.~M.}\
  \bibnamefont {Girvin}}, \bibinfo {author} {\bibfnamefont {L.}~\bibnamefont
  {Frunzio}}, \bibinfo {author} {\bibfnamefont {M.}~\bibnamefont {Mirrahimi}},
  \ and\ \bibinfo {author} {\bibfnamefont {M.~H.}\ \bibnamefont {Devoret}},\
  }\bibfield  {title} {{\selectlanguage {english}\enquote {\bibinfo {title}
  {{Autonomously stabilized entanglement between two superconducting quantum
  bits}},}\ }}\href {\doibase 10.1038/nature12802} {\bibfield  {journal}
  {\bibinfo  {journal} {Nature}\ }\textbf {\bibinfo {volume} {504}},\ \bibinfo
  {pages} {419--422} (\bibinfo {year} {2013})}\BibitemShut {NoStop}%
\bibitem [{\citenamefont {Leghtas}\ \emph {et~al.}(2013)\citenamefont
  {Leghtas}, \citenamefont {Vool}, \citenamefont {Shankar}, \citenamefont
  {Hatridge}, \citenamefont {Girvin}, \citenamefont {Devoret},\ and\
  \citenamefont {Mirrahimi}}]{leghtas_stabilizing_2013}%
  \BibitemOpen
  \bibfield  {author} {\bibinfo {author} {\bibfnamefont {Z.}~\bibnamefont
  {Leghtas}}, \bibinfo {author} {\bibfnamefont {U.}~\bibnamefont {Vool}},
  \bibinfo {author} {\bibfnamefont {S.}~\bibnamefont {Shankar}}, \bibinfo
  {author} {\bibfnamefont {M.}~\bibnamefont {Hatridge}}, \bibinfo {author}
  {\bibfnamefont {S.~M.}\ \bibnamefont {Girvin}}, \bibinfo {author}
  {\bibfnamefont {M.~H.}\ \bibnamefont {Devoret}}, \ and\ \bibinfo {author}
  {\bibfnamefont {M.}~\bibnamefont {Mirrahimi}},\ }\bibfield  {title} {\enquote
  {\bibinfo {title} {{Stabilizing a {Bell} state of two superconducting qubits
  by dissipation engineering}},}\ }\href {\doibase 10.1103/PhysRevA.88.023849}
  {\bibfield  {journal} {\bibinfo  {journal} {Phys. Rev. A}\ }\textbf {\bibinfo
  {volume} {88}},\ \bibinfo {pages} {023849} (\bibinfo {year}
  {2013})}\BibitemShut {NoStop}%
\bibitem [{\citenamefont {Reiter}\ \emph {et~al.}(2013)\citenamefont {Reiter},
  \citenamefont {Tornberg}, \citenamefont {Johansson},\ and\ \citenamefont
  {S{\o}rensen}}]{reiter_steady-state_2013}%
  \BibitemOpen
  \bibfield  {author} {\bibinfo {author} {\bibfnamefont {Florentin}\
  \bibnamefont {Reiter}}, \bibinfo {author} {\bibfnamefont {L.}~\bibnamefont
  {Tornberg}}, \bibinfo {author} {\bibfnamefont {G{\"o}ran}\ \bibnamefont
  {Johansson}}, \ and\ \bibinfo {author} {\bibfnamefont {Anders~S.}\
  \bibnamefont {S{\o}rensen}},\ }\bibfield  {title} {\enquote {\bibinfo {title}
  {{Steady-state entanglement of two superconducting qubits engineered by
  dissipation}},}\ }\href {\doibase 10.1103/PhysRevA.88.032317} {\bibfield
  {journal} {\bibinfo  {journal} {Phys. Rev. A}\ }\textbf {\bibinfo {volume}
  {88}},\ \bibinfo {pages} {032317} (\bibinfo {year} {2013})}\BibitemShut
  {NoStop}%
\bibitem [{\citenamefont {Huang}\ \emph {et~al.}(2013)\citenamefont {Huang},
  \citenamefont {Goan}, \citenamefont {Li},\ and\ \citenamefont
  {Milburn}}]{huang_generation_2013}%
  \BibitemOpen
  \bibfield  {author} {\bibinfo {author} {\bibfnamefont {Shang-Yu}\
  \bibnamefont {Huang}}, \bibinfo {author} {\bibfnamefont {Hsi-Sheng}\
  \bibnamefont {Goan}}, \bibinfo {author} {\bibfnamefont {Xin-Qi}\ \bibnamefont
  {Li}}, \ and\ \bibinfo {author} {\bibfnamefont {G.~J.}\ \bibnamefont
  {Milburn}},\ }\bibfield  {title} {\enquote {\bibinfo {title} {{Generation and
  stabilization of a three-qubit entangled \${W}\$ state in circuit {QED} via
  quantum feedback control}},}\ }\href {\doibase 10.1103/PhysRevA.88.062311}
  {\bibfield  {journal} {\bibinfo  {journal} {Phys. Rev. A}\ }\textbf {\bibinfo
  {volume} {88}},\ \bibinfo {pages} {062311} (\bibinfo {year}
  {2013})}\BibitemShut {NoStop}%
\bibitem [{\citenamefont {Aron}\ \emph {et~al.}(2014)\citenamefont {Aron},
  \citenamefont {Kulkarni},\ and\ \citenamefont
  {T{\"u}reci}}]{aron_steady-state_2014}%
  \BibitemOpen
  \bibfield  {author} {\bibinfo {author} {\bibfnamefont {Camille}\ \bibnamefont
  {Aron}}, \bibinfo {author} {\bibfnamefont {Manas}\ \bibnamefont {Kulkarni}},
  \ and\ \bibinfo {author} {\bibfnamefont {Hakan~E.}\ \bibnamefont
  {T{\"u}reci}},\ }\bibfield  {title} {\enquote {\bibinfo {title}
  {{Steady-state entanglement of spatially separated qubits via quantum bath
  engineering}},}\ }\href {\doibase 10.1103/PhysRevA.90.062305} {\bibfield
  {journal} {\bibinfo  {journal} {Phys. Rev. A}\ }\textbf {\bibinfo {volume}
  {90}},\ \bibinfo {pages} {062305} (\bibinfo {year} {2014})}\BibitemShut
  {NoStop}%
\bibitem [{\citenamefont {Underwood}\ \emph {et~al.}(2012)\citenamefont
  {Underwood}, \citenamefont {Shanks}, \citenamefont {Koch},\ and\
  \citenamefont {Houck}}]{underwood_low-disorder_2012}%
  \BibitemOpen
  \bibfield  {author} {\bibinfo {author} {\bibfnamefont {D.~L.}\ \bibnamefont
  {Underwood}}, \bibinfo {author} {\bibfnamefont {W.~E.}\ \bibnamefont
  {Shanks}}, \bibinfo {author} {\bibfnamefont {Jens}\ \bibnamefont {Koch}}, \
  and\ \bibinfo {author} {\bibfnamefont {A.~A.}\ \bibnamefont {Houck}},\
  }\bibfield  {title} {\enquote {\bibinfo {title} {{Low-disorder microwave
  cavity lattices for quantum simulation with photons}},}\ }\href {\doibase
  10.1103/PhysRevA.86.023837} {\bibfield  {journal} {\bibinfo  {journal} {Phys.
  Rev. A}\ }\textbf {\bibinfo {volume} {86}},\ \bibinfo {pages} {023837}
  (\bibinfo {year} {2012})}\BibitemShut {NoStop}%
\bibitem [{\citenamefont {Underwood}(2015)}]{underwood_thesis_2015}%
  \BibitemOpen
  \bibfield  {author} {\bibinfo {author} {\bibfnamefont {Devin~Lane}\
  \bibnamefont {Underwood}},\ }\href
  {http://gradworks.umi.com/36/86/3686679.html} {Ph.D. thesis},\ \bibinfo
  {school} {Princeton University} (\bibinfo {year} {2015})\BibitemShut
  {NoStop}%
\bibitem [{\citenamefont {Knap}\ \emph {et~al.}(2011)\citenamefont {Knap},
  \citenamefont {Arrigoni}, \citenamefont {von~der Linden},\ and\ \citenamefont
  {Cole}}]{knap_emission_2011}%
  \BibitemOpen
  \bibfield  {author} {\bibinfo {author} {\bibfnamefont {Michael}\ \bibnamefont
  {Knap}}, \bibinfo {author} {\bibfnamefont {Enrico}\ \bibnamefont {Arrigoni}},
  \bibinfo {author} {\bibfnamefont {Wolfgang}\ \bibnamefont {von~der Linden}},
  \ and\ \bibinfo {author} {\bibfnamefont {Jared~H.}\ \bibnamefont {Cole}},\
  }\bibfield  {title} {\enquote {\bibinfo {title} {{Emission characteristics of
  laser-driven dissipative coupled-cavity systems}},}\ }\href {\doibase
  10.1103/PhysRevA.83.023821} {\bibfield  {journal} {\bibinfo  {journal} {Phys.
  Rev. A}\ }\textbf {\bibinfo {volume} {83}},\ \bibinfo {pages} {023821}
  (\bibinfo {year} {2011})}\BibitemShut {NoStop}%
\bibitem [{\citenamefont {Nissen}\ \emph {et~al.}(2012)\citenamefont {Nissen},
  \citenamefont {Schmidt}, \citenamefont {Biondi}, \citenamefont {Blatter},
  \citenamefont {T{\"u}reci},\ and\ \citenamefont
  {Keeling}}]{nissen_nonequilibrium_2012}%
  \BibitemOpen
  \bibfield  {author} {\bibinfo {author} {\bibfnamefont {Felix}\ \bibnamefont
  {Nissen}}, \bibinfo {author} {\bibfnamefont {Sebastian}\ \bibnamefont
  {Schmidt}}, \bibinfo {author} {\bibfnamefont {Matteo}\ \bibnamefont
  {Biondi}}, \bibinfo {author} {\bibfnamefont {Gianni}\ \bibnamefont
  {Blatter}}, \bibinfo {author} {\bibfnamefont {Hakan~E.}\ \bibnamefont
  {T{\"u}reci}}, \ and\ \bibinfo {author} {\bibfnamefont {Jonathan}\
  \bibnamefont {Keeling}},\ }\bibfield  {title} {\enquote {\bibinfo {title}
  {{Nonequilibrium {Dynamics} of {Coupled} {Qubit}-{Cavity} {Arrays}}},}\
  }\href {\doibase 10.1103/PhysRevLett.108.233603} {\bibfield  {journal}
  {\bibinfo  {journal} {Phys. Rev. Lett.}\ }\textbf {\bibinfo {volume} {108}},\
  \bibinfo {pages} {233603} (\bibinfo {year} {2012})}\BibitemShut {NoStop}%
\bibitem [{\citenamefont {Grujic}\ \emph {et~al.}(2012)\citenamefont {Grujic},
  \citenamefont {Clark}, \citenamefont {Jaksch},\ and\ \citenamefont
  {Angelakis}}]{grujic_non-equilibrium_2012}%
  \BibitemOpen
  \bibfield  {author} {\bibinfo {author} {\bibfnamefont {T.}~\bibnamefont
  {Grujic}}, \bibinfo {author} {\bibfnamefont {S.~R.}\ \bibnamefont {Clark}},
  \bibinfo {author} {\bibfnamefont {D.}~\bibnamefont {Jaksch}}, \ and\ \bibinfo
  {author} {\bibfnamefont {D.~G.}\ \bibnamefont {Angelakis}},\ }\bibfield
  {title} {{\selectlanguage {english}\enquote {\bibinfo {title}
  {{Non-equilibrium many-body effects in driven nonlinear resonator arrays}},}\
  }}\href {\doibase 10.1088/1367-2630/14/10/103025} {\bibfield  {journal}
  {\bibinfo  {journal} {New Journal of Physics}\ }\textbf {\bibinfo {volume}
  {14}},\ \bibinfo {pages} {103025} (\bibinfo {year} {2012})}\BibitemShut
  {NoStop}%
\bibitem [{\citenamefont {Hur}\ \emph {et~al.}(2015)\citenamefont {Hur},
  \citenamefont {Henriet}, \citenamefont {Petrescu}, \citenamefont {Plekhanov},
  \citenamefont {Roux},\ and\ \citenamefont {Schir{\'o}}}]{hur_many-body_2015}%
  \BibitemOpen
  \bibfield  {author} {\bibinfo {author} {\bibfnamefont {Karyn~Le}\
  \bibnamefont {Hur}}, \bibinfo {author} {\bibfnamefont {Lo{\"i}c}\
  \bibnamefont {Henriet}}, \bibinfo {author} {\bibfnamefont {Alexandru}\
  \bibnamefont {Petrescu}}, \bibinfo {author} {\bibfnamefont {Kirill}\
  \bibnamefont {Plekhanov}}, \bibinfo {author} {\bibfnamefont {Guillaume}\
  \bibnamefont {Roux}}, \ and\ \bibinfo {author} {\bibfnamefont {Marco}\
  \bibnamefont {Schir{\'o}}},\ }\bibfield  {title} {\enquote {\bibinfo {title}
  {{Many-{Body} {Quantum} {Electrodynamics} {Networks}: {Non}-{Equilibrium}
  {Condensed} {Matter} {Physics} with {Light}}},}\ }\href
  {http://arxiv.org/abs/1505.00167} {\bibfield  {journal} {\bibinfo  {journal}
  {arXiv:1505.00167 [cond-mat, physics:quant-ph]}\ } (\bibinfo {year}
  {2015})},\ \bibinfo {note} {arXiv: 1505.00167}\BibitemShut {NoStop}%
\bibitem [{\citenamefont {Schwartz}\ \emph
  {et~al.}(2015{\natexlab{a}})\citenamefont {Schwartz}, \citenamefont {Martin},
  \citenamefont {Flurin}, \citenamefont {Aron}, \citenamefont {Kulkarni},
  \citenamefont {Tureci},\ and\ \citenamefont {Siddiqi}}]{ucbpaper}%
  \BibitemOpen
  \bibfield  {author} {\bibinfo {author} {\bibfnamefont {M.~E.}\ \bibnamefont
  {Schwartz}}, \bibinfo {author} {\bibfnamefont {L.}~\bibnamefont {Martin}},
  \bibinfo {author} {\bibfnamefont {E.}~\bibnamefont {Flurin}}, \bibinfo
  {author} {\bibfnamefont {C.}~\bibnamefont {Aron}}, \bibinfo {author}
  {\bibfnamefont {M.}~\bibnamefont {Kulkarni}}, \bibinfo {author}
  {\bibfnamefont {H.~E.}\ \bibnamefont {Tureci}}, \ and\ \bibinfo {author}
  {\bibfnamefont {I.}~\bibnamefont {Siddiqi}},\ }\href@noop {} {\bibfield
  {journal} {\bibinfo  {journal} {arXiv:1511.00702}\ } (\bibinfo {year}
  {2015}{\natexlab{a}})}\BibitemShut {NoStop}%
\bibitem [{\citenamefont {Boissonneault}\ \emph {et~al.}(2009)\citenamefont
  {Boissonneault}, \citenamefont {Gambetta},\ and\ \citenamefont
  {Blais}}]{boissonneault_dispersive_2009}%
  \BibitemOpen
  \bibfield  {author} {\bibinfo {author} {\bibfnamefont {Maxime}\ \bibnamefont
  {Boissonneault}}, \bibinfo {author} {\bibfnamefont {J.~M.}\ \bibnamefont
  {Gambetta}}, \ and\ \bibinfo {author} {\bibfnamefont {Alexandre}\
  \bibnamefont {Blais}},\ }\bibfield  {title} {\enquote {\bibinfo {title}
  {{Dispersive regime of circuit {QED}: {Photon}-dependent qubit dephasing and
  relaxation rates}},}\ }\href {\doibase 10.1103/PhysRevA.79.013819} {\bibfield
   {journal} {\bibinfo  {journal} {Phys. Rev. A}\ }\textbf {\bibinfo {volume}
  {79}},\ \bibinfo {pages} {013819} (\bibinfo {year} {2009})}\BibitemShut
  {NoStop}%
\bibitem [{\citenamefont {Schir{\'o}}\ \emph {et~al.}(2015)\citenamefont
  {Schir{\'o}}, \citenamefont {Joshi}, \citenamefont {Bordyuh}, \citenamefont
  {Fazio}, \citenamefont {Keeling},\ and\ \citenamefont
  {T{\"u}reci}}]{schiro_exotic_2015}%
  \BibitemOpen
  \bibfield  {author} {\bibinfo {author} {\bibfnamefont {M.}~\bibnamefont
  {Schir{\'o}}}, \bibinfo {author} {\bibfnamefont {C.}~\bibnamefont {Joshi}},
  \bibinfo {author} {\bibfnamefont {M.}~\bibnamefont {Bordyuh}}, \bibinfo
  {author} {\bibfnamefont {R.}~\bibnamefont {Fazio}}, \bibinfo {author}
  {\bibfnamefont {J.}~\bibnamefont {Keeling}}, \ and\ \bibinfo {author}
  {\bibfnamefont {H.~E.}\ \bibnamefont {T{\"u}reci}},\ }\bibfield  {title}
  {\enquote {\bibinfo {title} {{Exotic attractors of the non-equilibrium
  {Rabi}-{Hubbard} model}},}\ }\href {http://arxiv.org/abs/1503.04456}
  {\bibfield  {journal} {\bibinfo  {journal} {arXiv:1503.04456 [cond-mat,
  physics:quant-ph]}\ } (\bibinfo {year} {2015})},\ \bibinfo {note} {arXiv:
  1503.04456}\BibitemShut {NoStop}%
\bibitem [{afo(The coarse-graining cell in time can be made larger than
  $1/\kappa$)}]{afoot}%
  \BibitemOpen
  \href@noop {} {\  (\bibinfo {year} {The coarse-graining cell in time can be
  made larger than $1/\kappa$})}\BibitemShut {NoStop}%
\bibitem [{\citenamefont {Schoeller}\ and\ \citenamefont {Schon}(1994)}]{gsc}%
  \BibitemOpen
  \bibfield  {author} {\bibinfo {author} {\bibfnamefont {Herbert}\ \bibnamefont
  {Schoeller}}\ and\ \bibinfo {author} {\bibfnamefont {Gerd}\ \bibnamefont
  {Schon}},\ }\bibfield  {title} {\enquote {\bibinfo {title} {Mesoscopic
  quantum transport: Resonant tunneling in the presence of a strong coulomb
  interaction},}\ }\href@noop {} {\bibfield  {journal} {\bibinfo  {journal}
  {Phys. Rev. B}\ }\textbf {\bibinfo {volume} {50}},\ \bibinfo {pages} {18436}
  (\bibinfo {year} {1994})}\BibitemShut {NoStop}%
\bibitem [{\citenamefont {Mitra}\ \emph {et~al.}(2004)\citenamefont {Mitra},
  \citenamefont {Aleiner},\ and\ \citenamefont {Millis}}]{amitra}%
  \BibitemOpen
  \bibfield  {author} {\bibinfo {author} {\bibfnamefont {A.}~\bibnamefont
  {Mitra}}, \bibinfo {author} {\bibfnamefont {I.}~\bibnamefont {Aleiner}}, \
  and\ \bibinfo {author} {\bibfnamefont {A.~J.}\ \bibnamefont {Millis}},\
  }\bibfield  {title} {\enquote {\bibinfo {title} {Phonon effects in molecular
  transistors: quantal and classical treatment},}\ }\href@noop {} {\bibfield
  {journal} {\bibinfo  {journal} {Phys. Rev. B}\ }\textbf {\bibinfo {volume}
  {69}},\ \bibinfo {pages} {245302} (\bibinfo {year} {2004})}\BibitemShut
  {NoStop}%
\bibitem [{\citenamefont {Schwartz}\ \emph
  {et~al.}(2015{\natexlab{b}})\citenamefont {Schwartz}, \citenamefont {Martin},
  \citenamefont {Aron}, \citenamefont {Kulkarni},\ and\ \citenamefont
  {Tureci}}]{schwartz_toward_2015}%
  \BibitemOpen
  \bibfield  {author} {\bibinfo {author} {\bibfnamefont {M.~E.}\ \bibnamefont
  {Schwartz}}, \bibinfo {author} {\bibfnamefont {L.}~\bibnamefont {Martin}},
  \bibinfo {author} {\bibfnamefont {C.}~\bibnamefont {Aron}}, \bibinfo {author}
  {\bibfnamefont {M.}~\bibnamefont {Kulkarni}}, \ and\ \bibinfo {author}
  {\bibfnamefont {H.~E.}\ \bibnamefont {Tureci}},\ }\bibfield  {title}
  {\enquote {\bibinfo {title} {{Toward {Resource}-{Efficient} {Deterministic}
  {Entanglement} in 3D {Superconducting} {Qubits}}},}\ }\href
  {http://meetings.aps.org/link/BAPS.2015.MAR.A39.8} {\bibfield  {journal}
  {\bibinfo  {journal} {APS March Meeting}\ }\textbf {\bibinfo {volume}
  {A39.00008}} (\bibinfo {year} {2015}{\natexlab{b}})}\BibitemShut {NoStop}%
\bibitem [{\citenamefont {Malekakhlagh}\ and\ \citenamefont
  {Tureci}(2015)}]{malekakhlagh_origin_2015}%
  \BibitemOpen
  \bibfield  {author} {\bibinfo {author} {\bibfnamefont {Moein}\ \bibnamefont
  {Malekakhlagh}}\ and\ \bibinfo {author} {\bibfnamefont {Hakan~E.}\
  \bibnamefont {Tureci}},\ }\bibfield  {title} {\enquote {\bibinfo {title}
  {{Origin and {Implications} of \${A}{\textasciicircum}2\$-like {Contribution}
  in the {Quantization} of {Circuit}-{QED} {Systems}}},}\ }\href
  {http://arxiv.org/abs/1506.02773} {\bibfield  {journal} {\bibinfo  {journal}
  {arXiv:1506.02773 [cond-mat, physics:quant-ph]}\ } (\bibinfo {year}
  {2015})},\ \bibinfo {note} {arXiv: 1506.02773}\BibitemShut {NoStop}%
\bibitem [{\citenamefont {Schrieffer}\ and\ \citenamefont
  {Wolff}(1966)}]{schrieffer_relation_1966}%
  \BibitemOpen
  \bibfield  {author} {\bibinfo {author} {\bibfnamefont {J.~R.}\ \bibnamefont
  {Schrieffer}}\ and\ \bibinfo {author} {\bibfnamefont {P.~A.}\ \bibnamefont
  {Wolff}},\ }\bibfield  {title} {\enquote {\bibinfo {title} {{Relation between
  the {Anderson} and {Kondo} {Hamiltonians}}},}\ }\href {\doibase
  10.1103/PhysRev.149.491} {\bibfield  {journal} {\bibinfo  {journal} {Phys.
  Rev.}\ }\textbf {\bibinfo {volume} {149}},\ \bibinfo {pages} {491--492}
  (\bibinfo {year} {1966})}\BibitemShut {NoStop}%
\bibitem [{\citenamefont {Kraus}\ \emph {et~al.}(2008)\citenamefont {Kraus},
  \citenamefont {Buchler}, \citenamefont {Diehl}, \citenamefont {Kantian},
  \citenamefont {Micheli},\ and\ \citenamefont
  {Zoller}}]{kraus_preparation_2008}%
  \BibitemOpen
  \bibfield  {author} {\bibinfo {author} {\bibfnamefont {B.}~\bibnamefont
  {Kraus}}, \bibinfo {author} {\bibfnamefont {H.~P.}\ \bibnamefont {Buchler}},
  \bibinfo {author} {\bibfnamefont {S.}~\bibnamefont {Diehl}}, \bibinfo
  {author} {\bibfnamefont {A.}~\bibnamefont {Kantian}}, \bibinfo {author}
  {\bibfnamefont {A.}~\bibnamefont {Micheli}}, \ and\ \bibinfo {author}
  {\bibfnamefont {P.}~\bibnamefont {Zoller}},\ }\bibfield  {title} {\enquote
  {\bibinfo {title} {{Preparation of entangled states by quantum Markov
  processes}},}\ }\href@noop {} {\bibfield  {journal} {\bibinfo  {journal}
  {Phys. Rev. A}\ }\textbf {\bibinfo {volume} {78}},\ \bibinfo {pages} {042307}
  (\bibinfo {year} {2008})}\BibitemShut {NoStop}%
\bibitem [{\citenamefont {Diehl}\ \emph {et~al.}(2008)\citenamefont {Diehl},
  \citenamefont {Micheli}, \citenamefont {Kantian}, \citenamefont {Kraus},
  \citenamefont {B{\"u}chler},\ and\ \citenamefont
  {Zoller}}]{diehl_quantum_2008}%
  \BibitemOpen
  \bibfield  {author} {\bibinfo {author} {\bibfnamefont {S.}~\bibnamefont
  {Diehl}}, \bibinfo {author} {\bibfnamefont {A.}~\bibnamefont {Micheli}},
  \bibinfo {author} {\bibfnamefont {A.}~\bibnamefont {Kantian}}, \bibinfo
  {author} {\bibfnamefont {B.}~\bibnamefont {Kraus}}, \bibinfo {author}
  {\bibfnamefont {H.~P.}\ \bibnamefont {B{\"u}chler}}, \ and\ \bibinfo {author}
  {\bibfnamefont {P.}~\bibnamefont {Zoller}},\ }\bibfield  {title}
  {{\selectlanguage {english}\enquote {\bibinfo {title} {{Quantum states and
  phases in driven open quantum systems with cold atoms}},}\ }}\href@noop {}
  {\bibfield  {journal} {\bibinfo  {journal} {Nature Physics}\ }\textbf
  {\bibinfo {volume} {4}},\ \bibinfo {pages} {878--883} (\bibinfo {year}
  {2008})}\BibitemShut {NoStop}%
\bibitem [{\citenamefont {Cormick}\ \emph {et~al.}(2013)\citenamefont
  {Cormick}, \citenamefont {Bermudez}, \citenamefont {Huelga},\ and\
  \citenamefont {Plenio}}]{cormick_dissipative_2013}%
  \BibitemOpen
  \bibfield  {author} {\bibinfo {author} {\bibfnamefont {Cecilia}\ \bibnamefont
  {Cormick}}, \bibinfo {author} {\bibfnamefont {Alejandro}\ \bibnamefont
  {Bermudez}}, \bibinfo {author} {\bibfnamefont {Susana~F.}\ \bibnamefont
  {Huelga}}, \ and\ \bibinfo {author} {\bibfnamefont {Martin~B.}\ \bibnamefont
  {Plenio}},\ }\bibfield  {title} {{\selectlanguage {english}\enquote {\bibinfo
  {title} {{Dissipative ground-state preparation of a spin chain by a
  structured environment}},}\ }}\href {\doibase 10.1088/1367-2630/15/7/073027}
  {\bibfield  {journal} {\bibinfo  {journal} {New Journal of Physics}\ }\textbf
  {\bibinfo {volume} {15}},\ \bibinfo {pages} {073027} (\bibinfo {year}
  {2013})}\BibitemShut {NoStop}%
\bibitem [{\citenamefont {Lee}\ \emph {et~al.}(2013)\citenamefont {Lee},
  \citenamefont {Cho},\ and\ \citenamefont {Choi}}]{lee_emergence_2013}%
  \BibitemOpen
  \bibfield  {author} {\bibinfo {author} {\bibfnamefont {S.~K.}\ \bibnamefont
  {Lee}}, \bibinfo {author} {\bibfnamefont {J.}~\bibnamefont {Cho}}, \ and\
  \bibinfo {author} {\bibfnamefont {K.~S.}\ \bibnamefont {Choi}},\ }\bibfield
  {title} {\enquote {\bibinfo {title} {{Emergence of stationary many-body
  entanglement in driven-dissipative {Rydberg} lattice gases}},}\ }\href
  {http://arxiv.org/abs/1401.0028} {\bibfield  {journal} {\bibinfo  {journal}
  {arXiv:1401.0028 [quant-ph]}\ } (\bibinfo {year} {2013})},\ \bibinfo {note}
  {arXiv: 1401.0028}\BibitemShut {NoStop}%
\bibitem [{\citenamefont {Rao}\ and\ \citenamefont
  {M{\o}lmer}(2014)}]{rao_deterministic_2014}%
  \BibitemOpen
  \bibfield  {author} {\bibinfo {author} {\bibfnamefont {D.~D.~Bhaktavatsala}\
  \bibnamefont {Rao}}\ and\ \bibinfo {author} {\bibfnamefont {Klaus}\
  \bibnamefont {M{\o}lmer}},\ }\bibfield  {title} {\enquote {\bibinfo {title}
  {{Deterministic entanglement of {Rydberg} ensembles by engineered
  dissipation}},}\ }\href {http://arxiv.org/abs/1407.1228} {\bibfield
  {journal} {\bibinfo  {journal} {arXiv:1407.1228 [quant-ph]}\ } (\bibinfo
  {year} {2014})},\ \bibinfo {note} {arXiv: 1407.1228}\BibitemShut {NoStop}%
\bibitem [{\citenamefont {Reiter}\ \emph {et~al.}(2015)\citenamefont {Reiter},
  \citenamefont {Reeb},\ and\ \citenamefont
  {S{\o}rensen}}]{reiter_scalable_2015}%
  \BibitemOpen
  \bibfield  {author} {\bibinfo {author} {\bibfnamefont {Florentin}\
  \bibnamefont {Reiter}}, \bibinfo {author} {\bibfnamefont {David}\
  \bibnamefont {Reeb}}, \ and\ \bibinfo {author} {\bibfnamefont {Anders~S.}\
  \bibnamefont {S{\o}rensen}},\ }\bibfield  {title} {\enquote {\bibinfo {title}
  {{Scalable dissipative preparation of many-body entanglement}},}\ }\href
  {http://arxiv.org/abs/1501.06611} {\bibfield  {journal} {\bibinfo  {journal}
  {arXiv:1501.06611 [quant-ph]}\ } (\bibinfo {year} {2015})},\ \bibinfo {note}
  {arXiv: 1501.06611}\BibitemShut {NoStop}%
\bibitem [{\citenamefont {Rotondo}\ \emph {et~al.}(2015)\citenamefont
  {Rotondo}, \citenamefont {{Cosentino Lagomarsino}},\ and\ \citenamefont
  {Viola}}]{rotondo_dicke_2015}%
  \BibitemOpen
  \bibfield  {author} {\bibinfo {author} {\bibfnamefont {Pietro}\ \bibnamefont
  {Rotondo}}, \bibinfo {author} {\bibfnamefont {Marco}\ \bibnamefont
  {{Cosentino Lagomarsino}}}, \ and\ \bibinfo {author} {\bibfnamefont
  {Giovanni}\ \bibnamefont {Viola}},\ }\bibfield  {title} {\enquote {\bibinfo
  {title} {{Dicke {Simulators} with {Emergent} {Collective} {Quantum}
  {Computational} {Abilities}}},}\ }\href {\doibase
  10.1103/PhysRevLett.114.143601} {\bibfield  {journal} {\bibinfo  {journal}
  {Phys. Rev. Lett.}\ }\textbf {\bibinfo {volume} {114}},\ \bibinfo {pages}
  {143601} (\bibinfo {year} {2015})}\BibitemShut {NoStop}%
\bibitem [{\citenamefont {Verstraete}\ \emph {et~al.}(2009)\citenamefont
  {Verstraete}, \citenamefont {Wolf},\ and\ \citenamefont {{Ignacio
  Cirac}}}]{verstraete_quantum_2009}%
  \BibitemOpen
  \bibfield  {author} {\bibinfo {author} {\bibfnamefont {Frank}\ \bibnamefont
  {Verstraete}}, \bibinfo {author} {\bibfnamefont {Michael~M.}\ \bibnamefont
  {Wolf}}, \ and\ \bibinfo {author} {\bibfnamefont {J.}~\bibnamefont {{Ignacio
  Cirac}}},\ }\bibfield  {title} {{\selectlanguage {english}\enquote {\bibinfo
  {title} {{Quantum computation and quantum-state engineering driven by
  dissipation}},}\ }}\href@noop {} {\bibfield  {journal} {\bibinfo  {journal}
  {Nature Physics}\ }\textbf {\bibinfo {volume} {5}},\ \bibinfo {pages}
  {633--636} (\bibinfo {year} {2009})}\BibitemShut {NoStop}%
\bibitem [{\citenamefont {Hacohen-Gourgy}\ \emph {et~al.}(2015)\citenamefont
  {Hacohen-Gourgy}, \citenamefont {Ramasesh}, \citenamefont {De~Grandi},
  \citenamefont {Siddiqi},\ and\ \citenamefont
  {Girvin}}]{PhysRevLett.115.240501}%
  \BibitemOpen
  \bibfield  {author} {\bibinfo {author} {\bibfnamefont {S.}~\bibnamefont
  {Hacohen-Gourgy}}, \bibinfo {author} {\bibfnamefont {V.~V.}\ \bibnamefont
  {Ramasesh}}, \bibinfo {author} {\bibfnamefont {C.}~\bibnamefont {De~Grandi}},
  \bibinfo {author} {\bibfnamefont {I.}~\bibnamefont {Siddiqi}}, \ and\
  \bibinfo {author} {\bibfnamefont {S.~M.}\ \bibnamefont {Girvin}},\ }\bibfield
   {title} {\enquote {\bibinfo {title} {Cooling and autonomous feedback in a
  bose-hubbard chain with attractive interactions},}\ }\href@noop {} {\bibfield
   {journal} {\bibinfo  {journal} {Phys. Rev. Lett.}\ }\textbf {\bibinfo
  {volume} {115}},\ \bibinfo {pages} {240501} (\bibinfo {year}
  {2015})}\BibitemShut {NoStop}%
\bibitem [{\citenamefont {Agrawal}\ and\ \citenamefont
  {Pati}(2006)}]{agrawal_perfect_2006}%
  \BibitemOpen
  \bibfield  {author} {\bibinfo {author} {\bibfnamefont {Pankaj}\ \bibnamefont
  {Agrawal}}\ and\ \bibinfo {author} {\bibfnamefont {Arun}\ \bibnamefont
  {Pati}},\ }\bibfield  {title} {\enquote {\bibinfo {title} {Perfect
  teleportation and superdense coding with w states},}\ }\href {\doibase
  10.1103/PhysRevA.74.062320} {\bibfield  {journal} {\bibinfo  {journal} {Phys.
  Rev. A}\ }\textbf {\bibinfo {volume} {74}},\ \bibinfo {pages} {062320}
  (\bibinfo {year} {2006})}\BibitemShut {NoStop}%
\bibitem [{\citenamefont {Wang}\ \emph {et~al.}(2009)\citenamefont {Wang},
  \citenamefont {Yang}, \citenamefont {Su},\ and\ \citenamefont
  {Xie}}]{wang_simple_2009}%
  \BibitemOpen
  \bibfield  {author} {\bibinfo {author} {\bibfnamefont {Xin-Wen}\ \bibnamefont
  {Wang}}, \bibinfo {author} {\bibfnamefont {Guo-Jian}\ \bibnamefont {Yang}},
  \bibinfo {author} {\bibfnamefont {Yu-Huan}\ \bibnamefont {Su}}, \ and\
  \bibinfo {author} {\bibfnamefont {Min}\ \bibnamefont {Xie}},\ }\href@noop {}
  {\bibfield  {journal} {\bibinfo  {journal} {Quantum Inf Process}\ }\textbf
  {\bibinfo {volume} {8}} (\bibinfo {year} {2009})}\BibitemShut {NoStop}%
\bibitem [{\citenamefont {Leghtas}\ \emph {et~al.}(2015)\citenamefont
  {Leghtas}, \citenamefont {Touzard}, \citenamefont {Pop}, \citenamefont {Kou},
  \citenamefont {Vlastakis}, \citenamefont {Petrenko}, \citenamefont {Sliwa},
  \citenamefont {Narla}, \citenamefont {Shankar}, \citenamefont {Hatridge},
  \citenamefont {Reagor}, \citenamefont {Frunzio}, \citenamefont {Schoelkopf},
  \citenamefont {Mirrahimi},\ and\ \citenamefont
  {Devoret}}]{leghtas_confining_2015}%
  \BibitemOpen
  \bibfield  {author} {\bibinfo {author} {\bibfnamefont {Z.}~\bibnamefont
  {Leghtas}}, \bibinfo {author} {\bibfnamefont {S.}~\bibnamefont {Touzard}},
  \bibinfo {author} {\bibfnamefont {I.~M.}\ \bibnamefont {Pop}}, \bibinfo
  {author} {\bibfnamefont {A.}~\bibnamefont {Kou}}, \bibinfo {author}
  {\bibfnamefont {B.}~\bibnamefont {Vlastakis}}, \bibinfo {author}
  {\bibfnamefont {A.}~\bibnamefont {Petrenko}}, \bibinfo {author}
  {\bibfnamefont {K.~M.}\ \bibnamefont {Sliwa}}, \bibinfo {author}
  {\bibfnamefont {A.}~\bibnamefont {Narla}}, \bibinfo {author} {\bibfnamefont
  {S.}~\bibnamefont {Shankar}}, \bibinfo {author} {\bibfnamefont {M.~J.}\
  \bibnamefont {Hatridge}}, \bibinfo {author} {\bibfnamefont {M.}~\bibnamefont
  {Reagor}}, \bibinfo {author} {\bibfnamefont {L.}~\bibnamefont {Frunzio}},
  \bibinfo {author} {\bibfnamefont {R.~J.}\ \bibnamefont {Schoelkopf}},
  \bibinfo {author} {\bibfnamefont {M.}~\bibnamefont {Mirrahimi}}, \ and\
  \bibinfo {author} {\bibfnamefont {M.~H.}\ \bibnamefont {Devoret}},\
  }\bibfield  {title} {{\selectlanguage {english}\enquote {\bibinfo {title}
  {{Confining the state of light to a quantum manifold by engineered two-photon
  loss}},}\ }}\href {\doibase 10.1126/science.aaa2085} {\bibfield  {journal}
  {\bibinfo  {journal} {Science}\ }\textbf {\bibinfo {volume} {347}},\ \bibinfo
  {pages} {853--857} (\bibinfo {year} {2015})}\BibitemShut {NoStop}%
\end{thebibliography}%
\end{document}